\documentstyle[12pt,epsf,psfig]{article}
\def\bitem{\begin{itemize}}
\def\eitem{\end{itemize}}
\newcommand{\rr}{\bf r}
\newcommand{\pp}{\bf p}
\newcommand{\bb}{\bf b}
\newcommand{\KK}{($K^-,K^+$)\ }
\newcommand{\Km}{\mbox{$K^-$}\ }
\newcommand{\Kp}{\mbox{$K^+$}\ }

\newcommand{\mb}{{\rm mb}}

\newcommand{\srt}{\mbox{$\sqrt{s}$}}
\newcommand{\figposition}[1]{
\begin{center}
\begin{tabular}{c}
\hline
#1 \\
\hline
\end{tabular}
\end{center}}
\def\mathrm#1{{\rm #1}}

\def\nuc#1#2{\relax\ifmmode{}^{#1}{\protect\text{#2}}\else${}^{#1}$#2\fi}
\def\ack{\section*{Acknowledgement}%
  \addtocontents{toc}{\protect\vspace{6pt}}%
  \addcontentsline{toc}{section}{Acknowledgement}}
\newcommand{\TITLE}[3]{%
\begin{center}{\Large {#1}}\\ \vspace{0.5cm}{#2}\\ 
\vspace{0.3cm}{\it #3}\end{center}\vspace{0.3cm}}%

\begin{document}

\begin{titlepage}

\TITLE{
	$K^+$ momentum spectrum from \KK reactions\\
	in intranuclear cascade model
}{
	Y. Nara${}^a$\thanks{
		E-mail: ynara@nucl.phys.hokudai.ac.jp\ ,\ \ 
		Fax: +81-11-746-5444},
	A. Ohnishi${}^{a,b}$, T. Harada${}^c$ and A. Engel${}^d$
}{
	$a.$\ \ Department of Physics, Faculty of Science,
		Hokkaido University, Sapporo 060, Japan\\
	$b.$\ \ Nuclear Science Division, Lawrence Berkeley Laboratory,
		Berkeley, CA 94720, USA\\
	$c.$\ \ Department of Social Information,
		Sapporo Gakuin University, Ebetsu 069, Japan\\
	$d.$\ \ Department of Physics, Kyoto University,
		Kyoto, 606-01, Japan
}

\begin{abstract}
In a framework of intranuclear cascade (INC) type calculation,
  we study a momentum spectrum
  in reactions \KK at a beam momentum of 1.65 GeV/c.
INC model calculations are compared with
       the relativistic impulse approximation (RIA) calculations
  to perform the detailed study of the reaction mechanism.
We find that
	the INC model can reproduce the experimental data
	on various targets.
Especially,
	in the low-momentum region,
     the forward-angle cross sections of the $(K^-,K^+)$ reaction on
     from light to heavy targets are consistently
       explained with
	the two-step strangeness exchange and production processes
	with various intermediate mesons, 
       and
	$\phi$, $a_0$ and $f_0$ productions
        and their decay into $K^+K^-$.
In the two-step processes,
	inclusion of meson and hyperon resonances
	is found to be
	essential.
\end{abstract}

\end{titlepage}

\section{Introduction}

The study of nuclear systems
  with strangeness $S=$ --2
    is one of the most important and hottest subjects
    in nuclear physics,
  since
    these systems give us unique information on $YY$ interaction
     and
     they might be a doorway to study multi-strangeness systems
       such as the strange matter.
Until now, several ways to produce these $S=$ --2 nuclear systems 
           have been proposed~\cite{DW}.

The first one is the $\Xi^-$ absorption reaction at rest on nuclei.
In these thirty years, there are three reports on the discovery
of double-hypernuclei through the $\Xi^-$ absorption.
These events are found in the emulsion
by searching for the sequential weak decay processes
 (double stars)~\cite{stpxi2}.
Although the $\Xi^-$ absorption reaction
    is the most effective and the most direct way
    for the double-hypernucleus hunting,
  it is necessary to produce slow $\Xi^-$ particles, 
  to stop them in the emulsion,
  and to search for the double stars in the emulsion.
Thus the information is still too scarce to determine $YY$ interaction
at present, and further study along this line is necessary.
The second way to produce $S=-2$ nuclear systems
is the production of double-hypernuclei
at the $(K^-,K^+)$ reaction vertices on nuclei.
If $\Xi$-nucleus bound states exist and their widths are small enough,
we can see discrete peaks in $K^+$ momentum spectra
which directly reflect the $\Xi N$ interaction.
In addition to this clear signal, 
it is expected that double-hyperfragments produce
  after double-hyperon compound nuclear formation
  and evaporation process
  in the lower $K^+$ momentum region~\cite{Sano}.
Actually, KEK-E176~\cite{kkd} group
  have found $S=$ --2 nuclei at the $(K^-,K^+)$ 
  reaction point in the emulsion target,
  although its mass cannot be specified.
The $(K^-,K^+)$ reaction also provides us with $\Xi^-$ particles which 
can be used as a source of $\Xi^-$ absorption at rest.
Therefore, the study of the $(K^-, K^+)$ reaction on nuclei
is important and urgent to explore $S=$ --2 nuclear systems.
The relativistic heavy-ion collision is 
considered to be the third way to produce $S=$ --2 nuclear systems.
In this reaction, more strange nuclei ($S\leq -3$) or strangelets
are expected to be produced.
However, at the current stage, there is no positive report on 
the discovery of these nuclei,
probably because the production cross section is too small to observe.

Recently,
it has become experimentally possible
   to study double strangeness exchange reactions
            \KK on nuclear targets.
For example, small angle \KK cross sections
 at $p_{K^-}=1.65$ GeV/c were
 measured on several targets
 at KEK-PS
 by Iijima {\it et al.}~\cite{Iijima}.
The measured $K^+$ momentum spectrum shows
  a striking structure
     which gives rise to a target mass number dependence ($A$-dependence):
     In addition to quasifree peaks which correspond to 
     elementary process of $K^-p \to K^+\Xi^-$,
     there appear a large bump spreading
		 from $p_{K^+}=0.35$ GeV/c to $1.0$ GeV/c, 
  and the yield of this bump in the lower momentum region
     grows much faster $(\sim A^{0.6})$
     than that of the quasifree peak $(\sim A^{0.3})$
       as the target mass increases.
The comparison of the measured spectrum
  with the DWIA spectrum indicates that
    the high momentum peak $p_{K^+}\approx$ 1.1 GeV/c
       has been understood by
    quasifree process $K^- p \to K^+ \Xi$.
However, the huge lower momentum bump cannot be explained
	by the quasifree processes only with well-known elementary reactions
	such as $K^- p \to K^+ \Xi^*$~\cite{Iijima}.
Thus, the appearance of this lower peak bump has casted a doubt
on the belief that quasifree processes dominate in meson induced reactions
 at around 1 GeV/c region,
 and at the same time,
it gives us a hope that we can make more abundant double-hypernuclei
from the unknown mechanism in this large bump.
Therefore, it is important to obtain the knowledge for
  the reaction mechanism of the \KK reaction on nuclei.

Recently, one of possible mechanisms
 is proposed by Gobbi {\it et al.}~\cite{DG}.
They have claimed that
  a large part of the lower momentum peak
  is exhausted by
  the contribution from the heavy-meson production followed by its decay:
	$K^- p  \to  M \Lambda, \  M  \to K^+ K^-,$
   where $M$ is the $f_0(975)$, $a_0(980)$ or $\phi(1020)$ meson.
It should be noted that the first reaction is 
a so-called subthreshold particle production.
The treatment of the nucleon Fermi motion is considered to be important.
Since the incident $K^-$ particle
 has a large cross section with a nucleon in nuclear targets, 
the first-step reaction occurs mainly in the surface region 
where the Fermi momentum is smaller than that in the inner region.
Although this reduction of the effective Fermi momentum
 should largely suppress
 the subthreshold particle production,
it is not taken into account by Gobbi {\it et al.}
In fact, we can see much less kaon pair
creations $M\to K^+K^-$ experimentally 
in the scintilator-fiber detector~\cite{Imai-Private}
than those expected in their calculations.
Therefore, more detailed analyses of these reactions are desired
before we conclude that the lower momentum bump comes from 
heavy-meson decays.

Another plausible process is
  the two-step strangeness exchange and production reaction,
  such as $K^-p \to M'Y$ followed by $M'N\to K^+Y$,
  where $M'$ and $Y$ denote intermediate mesons
  $(\pi, \rho, \eta, \ldots)$ and hyperons
  $(\Lambda, \Sigma, \Lambda^*, \Sigma^*)$, respectively.
In previous works~\cite{Iijima,DG}, the contributions of 
these two-step processes are estimated to be too small to explain 
the lower-momentum huge bump.
However, these estimations limit the intermediate mesons and
hyperons to pions and ground state hyperons, respectively. 
In addition, they considered only the \nuc{12}{C} target case.
As pointed out by Iijima {\it et al.}~\cite{Iijima},
     other possible reactions may contribute to the 
	$K^+$ spectrum significantly.
Furthermore,
      the target mass number dependence
      in the experimental spectrum
      is well-fitted with $A^{0.56\pm 0.02}$
      in the low momentum region,
    while
      $A^{0.38\pm 0.03}$
      in the high momentum region.
This scaling property strongly suggests
    the contribution of multi-step processes~\cite{Iijima}.
Therefore,
  it is necessary to treat multi-step processes correctly
  to evaluate the $K^+$ momentum spectrum.

In this paper,
we investigate multi-step effects
 in the \KK reactions on the various targets,
 i.e., \nuc{12}{C}, \nuc{27}{Al}, \nuc{63}{Cu}, \nuc{107}{Ag}
      and \nuc{208}{Pb},
 using the intranuclear cascade (INC) model~\cite{INC1,INC2,INC3,INC4}.
One of our aims is
          to see the target mass number dependence on
    the \Kp momentum distribution in the \KK reaction.
Thus as a first step,
  we neglect some other effects
    such as the influence of the potential effect
    between hadrons.
We will also compare the results of the INC spectrum
 with the one-step process spectrum in the
 relativistic impulse approximation (RIA) model~\cite{Horowitz}
 to study the reaction $(K^-,K^+)$ in detail.

In order to calculate the contributions of various two-step processes, 
there are two problems to be settled.
The first problem is that they involve several elementary processes
such as $\rho N \to KY$, 
whose cross sections cannot be measured experimentally.
In this paper,
  we adopt the generalized Breit-Wigner formula~\cite{Brown,Sorge},
in which $s$ channel dominance is assumed and the interference effects 
between different baryon resonances are ignored.
It has been shown that this formula works well in the energy range 
under consideration, and it reduces theoretical ambiguities.
The second problem lies in the treatment of complex multi-step processes.
For the discussion of multi-step processes,
    cascade type models
    are considered to be the most reliable ones
    at the current stage:
  Each elementary two-body collision process is treated explicitly,
	then it is easy to include various multi-step processes.
The INC model
	does quite well for many situations.
After the first application 
	to high-energy nuclear collisions~\cite{INC1,INC2,INC3,INC4},
	it is also applied to pion-nucleus scattering~\cite{Pion1,Pion2}.
In these ten years, considerable efforts have been devoted to extend
	the INC model.
One of them is the inclusion of mean-field effects
	within one-body theories
	~\cite{BUU1,BUU2,BUU3,BUU4,BUU5,VUU1,VUU2,VUU3,VUU4}
	and $N$-body dynamics~\cite{QMD1,QMD2,QMD3,QMD4,Maru1,Maru2,nQMD}.
Furthermore,
	it becomes possible to describe the time-development of 
	totally anti-symmetrized wave functions~\cite{HORIUCHI,AMD2}.
These extended versions of the INC model are
	called microscopic transport models or microscopic simulations,
	and they have been successful in studying various aspects of 
	heavy-ion collisions as well as light-ion induced reactions
	---
	particle production~\cite{mRQMD1,mRQMD2,Mosel,BUU4},
	single-particle spectra~\cite{VUU2},
	collective flow~\cite{BUU1,AMD2},
	and fragment production~\cite{QMD2,Maru1,AMD,StoppedK}.

This paper is organized as follows:
We describe our INC model in Sec.~\ref{sec:inc},
 and present a RIA model in Sec.~\ref{sec:ria}.
In Sec.~\ref{sec:cross},
  elementary cross sections 
    are presented.
In Sec.~\ref{sec:results}, 
  we analyze the experimental data at KEK-E176~\cite{Iijima}
  and discuss the reaction mechanism of the $(K^-,K^+)$ reactions.
Finally, Section~\ref{sec:sumkk} contains our conclusion.

\section{Intranuclear Cascade (INC) model}\label{sec:inc}

The Intranuclear cascade (INC) model we adopt here describes
	the propagation of the leading particle
	(the most energetic mesons at each time step).
The momentum distribution of a nucleon in the target
	is assumed to be that of the Fermi gas,
	and the probability and the point of the collision
	between leading particle and the nucleon
	are determined by
	the mean free path of the propagating leading particle.
This prescription is not fully microscopic,
	since the dynamical evolution of other particles is ignored.
For example,
	we ignore the change of the nucleon momentum distribution
	and the secondary collision of a produced hyperon.
However,
	this model is enough to describe the $K^+$ meson production,
	because other particles than the leading particle
	cannot produce the $K^+$ meson energetically
	and distortion of the target density is expected to be
	only important after the $K^+$ meson go out from the target.

Most of cascade type models are realized 
	by the Monte-Calro procedure.
There are several different prescriptions for
	a cascading of the colliding particles.
One method is 
	a so-called ``closest distance approach"~\cite{INC3,INC4};
if the minimum relative distance $r_{min}$ for any pair of particles
	becomes less than $\sqrt{\sigma(\srt)/\pi}$,
	that pair is assumed to collide with each other,
	where $\sigma(\srt)$ is the total cross section
	for the pair at the c.m. energy $\srt$.
In another one,
	the collision frequency is determined by the mean free path.
For example,
	the collision will take place with the probability
	$\sigma(\srt)v\rho({\rr}) dt$~\cite{INC1,INC2,Pion2},
	where $v$ and $\rho({\rr})$ are the relative velocity
	of the colliding pair and the nuclear density, respectively.
We have checked that the both methods give the same results
	for the $K^+$ momentum distribution in the $(K^-,K^+)$ reaction
However, 
	other details may be different in several computer codes,
	therefore we explain our Monte-Calro procedure
	for completeness as follows:

\begin{itemize}
\item[1.]
The incoming \Km meson which is the first leading particle
	is boosted at the incident momentum $p_{K^-}=1.65$GeV/c
	in the laboratory frame.
\item[2.]
To determine where this particle collides with a nucleon
	in the target nucleus \nuc{A}{Z},
	we generate the Fermi momentum randomly
	with $p_F=(3\pi^2\rho_1)^{1/3}$,
	where $\rho_1=\rho Z/A$, $\rho=Z/(4/3 \pi R^2)$
	and $R=1.2A^{1/3} {\rm (fm)}$ in the case of matter distribution
	for the nucleus.
	If we employ the Fermi type nuclear density,
	the local Tomas-Fermi approximation is used to
	obtain the Fermi momentum $p_F({\rr})$.
\item[3.]
The total cross section
	between the leading particle and
	the proton(neutron) $\sigma_{\mathrm{p}}$($\sigma_{\mathrm{n}}$)
	is determined at this energy. 
	If the leading particle is a resonance,
	the decay probability per unit length is calculated as
	$P_d=\mit\Gamma/(v\gamma)$,
	where $\mit\Gamma$ is the total width of the resonance,
	$v$  and $\gamma$ denote the velocity and Lorentz gamma factor,
	respectively.
\item[4.]
The mean free path is obtained as
	 $\lambda = 1/(P_p+P_n+P_d)$,
           where $P_{\mathrm{p}}=\sigma_{\mathrm{p}}\rho_{\mathrm{p}}$,
		 $P_{\mathrm{n}}=\sigma_{\mathrm{n}}\rho_{\mathrm{n}}$.
	We determine randomly the collision point
	according to the probability distribution
	$dP/\Delta r=\exp(-\Delta r/\lambda)/\lambda$:
	This distribution is realized by choosing $\Delta r$ as follows.
         \begin{equation}
            {\rr}'= {\rr} + \Delta r{{\pp} \over |{\pp}|}
         \end{equation}
	 with $\pp$ is the momentum of the leading particle,
         where
                $\Delta r=-\lambda \log(1-x)$
            and $x$ is a random number between $0 \leq x<1$.
\item[5.]
If $|{\rr}'|$ is larger than the nuclear radius $R$,
	cascading is regarded as finished.
	Otherwise, we select the branch 
	randomly according to the probabilities of
	$P_{\mathrm{p}}$, $P_{\mathrm{n}}$ and $P_{\mathrm{d}}$.
\item[6.]
If the leading particle is determined to
	collide with other nucleon,
       	the collision branch is chosen randomly
	according to the partial cross sections,
	i.e. the elastic
       	or inelastic collision
	between the leading particle and the nucleon.
	After choosing the branch,
	the kinematics of the collision is calculated,
	and the final momenta of particles after this collision
	are determined.
\item[7.]
If particles after the collision are nucleons,
	the Pauli blocking is checked.
	If its momentum in the laboratory frame 
	is smaller than the Fermi momentum 270 MeV/c,
	this collision is considered to be Pauli blocked
	and nothing occurs.
\item[8.]
	The procedures from 2 to 7 are repeated
	until this event finishes.
\end{itemize}

It is useful to mention here that
	our cascade procedure can be used
	only for the matter (constant) nuclear density
	$\rho=\rho_0\theta(R-|{\rr}|)$.
If more realistic density is used in the calculations,
	as the Woods-Saxon type density,
	the collision prescription has to be modified.
In that case,
	we change the collision rate according to Ref.~\cite{INC1,Pion2}
	or ``closest distance approach".
We have checked that the calculated $K^+$ momentum distribution
	does not depend on the cascading prescription.

Since
	the total cross sections are very small in the \KK reaction, 
	note that a large amount of computational time is required
	to get sufficient statistics
	within the Monte Calro technique.
In this work, therefore,
	while all the elastic scatterings and
	the relevant inelastic reactions described in Sec.\ref{sec:cross}
	are included,
	we ignore other non-interesting inelastic processes
	which do not result in the $K^+$ production;
	the probabilities for relevant inelastic reactions are
	increased by the factor of
\begin{equation}
  F = {\sigma_{tot}-\sigma_{el} \over \sum\sigma_k}\ ,
\end{equation}
where  $\sigma_k$ denotes the partial cross section
	for relevant inelastic reactions,
	and $\sigma_{tot}$ and $\sigma_{el}$ are
	the total and elastic cross sections
	at each collision energy, respectively.
This procedure modifies the weight
	for each simulation,
	since we only treat the elastic scattering and
	relevant reactions.
Thus 
	when the relevant inelastic branch is chosen,
	the weight for the $j$-th simulation event
	after $i$-th step is defined as
\begin{equation}
  w_i^j=w_{i-1}^j{1\over F}\ ,
\end{equation}
where 
	the initial value is $w^j_0=1$.
Under this assumption,
	the probability to detect the $K^+$ meson
	with a given impact parameter $b$
	is calculated as
\begin{equation}
{d^2 P(b) \over dp d\Omega}={\sum_{j} w^j \Theta^j \over N_{ev}}
                            {1 \over d\Omega_d dp}
\end{equation}
with
	$d\Omega_d = 2\pi (\cos\theta_{min}-\cos\theta_{max})$,
	where $\theta_{min}=1.7^{\circ}$ and $\theta_{max}=13.6^{\circ}$
	are the minimum and the maximum angles
       	detected by the experimental measurement~\cite{Iijima},
	respectively,
	and $N_{ev}$ represents
	the number of events generated in the simulation.
The weight $w^j$ is
  defined as the final weight of the $j$-th event,
	and $\Theta^j=1$ if the $K^+$ particle enters the detector range,
	otherwise $\Theta^j=0$.
Typically, we generate
	1,000,000 events in this calculations.

In order to compare the theoretical results
	with the experimental data~\cite{Iijima},
	we integrate the probability over the impact parameter ${\bb}$:
\begin{equation}
\left\langle
  {d^2 \sigma \over dp d\Omega}
\right\rangle^f
  =\int d{\bb} {d^2 P(b) \over dp d\Omega},
\end{equation}
where  the definition of
$\left\langle
\right\rangle^f$
is the forward cross section
	averaged over the detected solid angle in the laboratory frame.

\section{Relativistic impulse approximation (RIA) model}
\label{sec:ria}

\def\MSquare{\langle \left| {\cal M} \right|^2 \rangle_{S}} 

In this section,
	we explain
	the relativistic impulse approximation (RIA)~\cite{Horowitz}
	whose results are compared with those of INC.
The RIA calculation in this paper
	is the standard one in the quasifree region,
	and is essentially the same
	as Iijima {\it et al.}~\cite{Iijima}.
The one-step cross section
	$K^- + p \to K^+ + \Xi^- (\Xi^{*-})$
	is calculated from
	the invariant matrix element ${\cal M}$ 
	and the momentum distribution of a nucleon
	in the target nucleus:
\begin{eqnarray}
\label{riaA}
d\sigma &=&
  Z_{eff}(K^-,K^+)\
  \int {d^3p_2 d{\mit\Gamma_3} d{\mit\Gamma_4} \over 4 E_1 E_2 v_{12}}\ 
  f(p_2)\ 
  \MSquare\ 
  (2\pi\hbar)^4 \delta^4(p_1+p_2-p_3-p_4)\ ,\\ 
d\mit\Gamma_i
  &=&
   {d^3p_i \over 2E_i (2\pi\hbar)^3}\ , \quad
   f(p)	=
   \left({4\pi P_F^3 \over 3}\right)^{-1}\ \theta(P_F-p) \label{riaB} \ ,
\end{eqnarray}
where
  the suffixes 1, 2, 3 and 4 represent
  $K^-, p, K^+$ and $\Xi^-$, respectively.
As for the proton momentum distribution in nuclei $f(p_2)$, 
  we have adopted Fermi gas model with $P_F = 0.27$ GeV/c.
In this calculation,
 we use the parameterized elementary differential cross sections,
 as we will express in Sec.~\ref{sec:cross},
 so as to calculate
 the spin-averaged invariant matrix element squared $\MSquare$
 for elementary process of
      $K^-+p \to K^+ + \Xi^-/\Xi^{*-}(1530)$.
The elementary cross section is written as
\begin{equation}
{d\sigma^{elem.}(s) \over d\Omega_{CM}} 
	= {1 \over 4 E_1 E_2 v_{12}}\ {1 \over (2\pi)^2}\ 
		{p_f \over 4 \sqrt{s}}\ 
		\MSquare\ ,
\label{riaC}
\end{equation}
where
  $\sqrt{s}$ and $p_f$ are the invariant mass
      and the final c.m. momentum
      of the elementary process, respectively.
Using Eqs.~(\ref{riaA}) (\ref{riaB}) (\ref{riaC}),
  we get the one-step quasifree cross sections
  as follows,
\begin{equation}
\label{ria-dirB}
{d^2\sigma \over dp_3 d\Omega_3}
	=
	Z_{eff}(K^-,K^+)\
	{p_3^2 \over E_3}\ 
	\int {d^3p_2 d^3p_4 \over E_4}\ 
		f(p_2)\ 
		{\sqrt{s} \over p_f}\ 
		{d\sigma^{elem.}(s) \over d\Omega_{CM}}\ 
		\delta^4(p_1+p_2-p_3-p_4) \ . 
\end{equation}
In the RIA model,
  distortions of the initial $K^-$ and final $K^+$ mesons
     are usually treated in the eikonal approximation.
Thus the effective proton number $Z_{eff}(K^-,K^+)$ which
      represents the collision frequency of the particles,
      is given by
\begin{equation}
\label{zeff}
 Z_{eff}(K^-,K^+) = {Z\over A}\int d{\rr} \rho({\rr})\exp\left[
	 -\sigma_{K^-}\int^z_{-\infty}\rho(x,y,z')dz'
	 -\sigma_{K^+}\int_z^{\infty}\rho(x,y,z')dz'
	   \right] \ ,
\end{equation}
where $\rho({\rr})$ denotes the nuclear density,
	and $\sigma_{K^-}=29.0$mb and $\sigma_{K^+}=18.4$mb are
	\Km-nucleon and \Kp-nucleon total cross sections, respectively.

For productions of
the intermediate vector and scalar mesons and their decay,
we use following expressions in RIA:
\begin{eqnarray}
\nonumber
d\sigma
	&=&
	Z_{eff}(K^-,M)\
	\int {d^3p_2 d{\mit\Gamma_4} \over 4 E_1 E_2 v_{12}}\ 
		{d\omega_3 d^3p_3 \over (2\pi\hbar)^4}\ 
		f(p_2)\ 
		\MSquare\ 
		(2\pi\hbar)^4 \delta^4(p_1+p_2-p_3-p_4) \\
\label{ria-decA}
	&&
		\phantom{MMMM}\times
		d\mit\Gamma_5  d\mit\Gamma_6 \ 
		(2\pi\hbar)^4 \delta^4(p_3-p_5-p_6) \\
{\cal M}
	&=& 	{\cal M}_{12\to34}\ 
		G_3(\omega_3, {\bf p}_3)\ 
		{\cal M}_{3\to56}
\end{eqnarray}
where the four momentum of the meson is represented as 
$p_3 = (\omega_3, {\bf p}_3)$,
and the suffixes 1, 2, 3, 4, 5 and 6 represent
$K^-, p, M, \Lambda, K^+$ and $K^-$, respectively.
In this reaction, we have assumed 
  that (1) the lifetime of the meson is long enough,
  and (2) the decay occurs isotropically.
The assumption (1) is realized 
	by the approximation to the Green function $G_3$.
In the case of the scalar meson, the Green function is given by
\begin{eqnarray}
\left| G_3 (\omega_3, {\bf p}_3) \right|^{-2}
   &=&
	\left| \omega_3^2 - E_3^2 + iM_3 \mit\Gamma_3 \right|^2
  \simeq
 {2 \pi \over 4 E_3 M_3 \mit\Gamma_3 }\ \delta(\omega_3 - E_3)\ ,\\
E_3	&=&	\sqrt{M_3^2 + {\bf p}_3^2}\ ,
\end{eqnarray}
where $\mit\Gamma_3$ is the total decay width of the meson.
Note that the assumption (1) is already used
	in Eq.(\ref{ria-decA}) implicitly,
	since the effective proton number $Z_{eff}(K^-,M)$ is
	calculated under the assumption that the meson $M$ does not
	decay inside the nucleus.
The matrix element ${\cal M}_{3\to56}$
     can be calculated
   under the assumption (2), which is obtained as
\begin{equation}
\left|{\cal M}_{3\to56}\right| ^2 
	= {8\pi M_3^2 \mit\Gamma_{3\to56} \over p_f}\ , \quad
p_f 	= \sqrt{M_3^2 - 4 M_5^2}/2\ .
\end{equation}
Thus,
   the differential cross section for the $K^+$ particle
       through the meson decay can be expressed as
\begin{eqnarray}
&& E_{K^+} {d^3\sigma \over dp_{K^+}^3}
	=
	\left( {M_M \over M_{K}} \right)^2 \ 
	Br(M \to K^+K^-)
	\int {d\Omega_M \over 4\pi}\ 
	E_3\ {d^3\sigma(K^-A \to M\Lambda) \over dp_M^3}\ , \\
&& Br(M \to K^+K^-)
	=
		\mit\Gamma_{M \to K^+K^-}/\mit\Gamma_M \ ,
\end{eqnarray}
where the solid angle $\Omega_M$
	is calculated in the rest frame of $K^+$, 
	and
	$E_3d^3\sigma(K^-A \to M\Lambda) / dp_M^3$
	denotes the invariant cross section
	for the one-step intermediate meson production per a proton, 
	which is given by Eq.(\ref{ria-dirB})
	without the effective proton number $Z_{eff}$.

\section{ Elementary cross sections}\label{sec:cross}

Now let us explain the elementary cross sections used in this paper.
 All the energies and cross sections
  are give in GeV and mb unit, respectively.
We have categorized the events into three types.
The first category is the direct-type reaction,
	which concerns the one-step double strangeness exchange
	reaction $({\bar K} N \to K \Xi, K\Xi^*)$.
The second one is the heavy scalar and vector meson productions
	and their decay $({\bar K} N \to M Y, M \to K^-K^+)$
	proposed by Gobbi {\it et al.}~\cite{DG}.
The third type is
  the two-step strangeness exchange and production reaction,
  such as ${\bar K} N \to \pi Y$ followed by $\pi N \to K Y$.
We will explain the necessary elementary cross sections in this order.

The following processes are necessary to describe the direct-type
	double strangeness exchange reaction.
\begin{itemize}
\item[(A)]
	All the elastic scatterings and
	charge exchange reactions of $KN$ and ${\bar K}N$.
\item[(B)]
	Double strangeness exchange processes:
	\bitem
	\item[]
		${\bar K} N \to K \Xi,\  K \Xi^*(1530).$
	\eitem
\item[(C)]
	Initial anti-kaon interaction:
	\bitem
	\item[]
		${\bar K} N \to {\bar K}^*(892) N$.
	\eitem
\item[(D)]
	Decay of resonance kaons (anti-kaons):\
	\bitem
	\item[]
		$K^*(892) \to K \pi$,\
		${\bar K}^*(892) \to {\bar K} \pi$,\ 
	\eitem
\item[(E)]
	Final kaon interactions:\
	\bitem
	\item[]
	  $K N \to (K^*(892) N ,\ K \Delta(1232)$),
	\item[]
          $K^*(892)N  \to (K N, \ K^*(892) \Delta(1232)$).
	\eitem
\end{itemize}
Although the actual strangeness exchange occurs
  only in the processes of (B), 
  initial and final state interactions have to be included
when we are interested in the absolute and whole spectrum of $K^+$
including the lower momentum region,
as will be demonstrated in Sec.~\ref{sec:results}.

The above cross sections
   are fitted to the experimental data~\cite{HERA1,HERA2}
	with appropriate functions or tabulated.
The total, elastic and charge exchange cross sections of
  (A) are well-known experimentally~\cite{HERA1,PG}.
We employ here the smooth interpolations of these data.
Angular dependencies of the elastic collisions are
	simply assumed to have the form~\cite{INC3,BUU1}
\begin{equation}
  { d\sigma \over dt } \propto \exp({b(\srt)t})\ ,
\end{equation}
where
   $t$ denotes the square of the transferred four momentum,
   \srt\  is the c.m. energy,
   and $b(\srt)$ is taken to be
       $b(\srt)=0.125\sigma(\srt)_{tot}$.

The double strangeness exchange reactions (B) are 
	also well-known experimentally
	although the errorbars are still large.
Therefore, we use following parameterization
 for the elementary $\Xi$ and $\Xi^*(1530)$ production cross sections:
\begin{equation}
\sigma(K^- p \to K^+ \Xi^-) = \left\{
  \begin{array}{ll}
   \displaystyle{
  {0.0413463(\srt - 1.81496)^{1.68792}
	\over (\srt - 1.93536)^2 + 0.01174 }}
  \quad [\mb]  & \qquad \srt \leq 2.22 \mathrm{[GeV/c]}\\[0.2cm]
   \displaystyle{
      {0.00613(\srt - 1.81496)^{0.48020}
	    \over (\srt - 2.02435)^2+0.01739}
   } & \\
   \displaystyle{
     +{0.001673(\srt - 2.11496)^{0.338065}
		  \over(\srt - 2.3174)^2+0.016429}
   } \quad [\mb]
    & \qquad \srt > 2.22 \mathrm{[GeV/c]}
  \end{array}\right.
\end{equation}
\begin{eqnarray}
\sigma(K^- p \to K^0 \Xi^0) & = &
  {0.00364308(\srt - 1.812571)^{0.620511}
	  \over (\srt - 2.05469)^2 + 0.0184015}
  \quad [\mb]\ , \\
\sigma(K^- p \to K^+ \Xi^{*-}(1530)) & = &
  {0.0017692(\srt - 2.0264)^{0.45434}
	    \over (\srt - 2.09274)^2 + 0.01394}
  \quad [\mb]\ , \\
\sigma(K^- p \to K^0 \Xi^{*0}(1530)) & = &
  {0.003854(\srt - 2.0264)^{0.3475}
	  \over (\srt - 2.18855)^2 + 0.028487}
  \quad [\mb]\ .
\end{eqnarray}
These fitted cross sections
   are shown in Fig.~\ref{CrossXi},
   together with the experimental data.
Although we can take another parameterizations
    within the experimental error bars,
 we have checked that
    our simulation results are almost the same.
\figposition{Fig.~\ref{CrossXi}}
Around the incident momentum of $p_{K^-}=1.65$ GeV/c for the $K^-$ beam,
  the experimental data
  shows a backward peak angular distribution.
We fit angular distributions of $\Xi$ production cross sections
  to the experimental data at $p_{K^-}=1.7$ GeV/c,
       neglecting these energy dependence:
\begin{equation}
{d\sigma \over d\Omega}  \sim \left\{
  \begin{array}{lr}
    0.34308\exp(0.6229\cos\theta) + 0.06681\exp(-4.4396\cos\theta) 
    & \quad {\rm for }\ K^-p \to K^+ \Xi^-\ ,\\
    0.44\exp(3.1684\cos\theta) + 0.5428\exp(-1.6747\cos\theta)
    & \quad {\rm for }\ K^-p \to K^0 \Xi^0\ .
  \end{array}\right.
\end{equation}
As shown in Fig.~\ref{AngXi},
  this parameterization is a good approximation
  within a Fermi momentum spreading around $p_{K^-}=1.65$ GeV/c.
For the $\Xi^*(1350)$ production,
  the angular distributions 
    are assumed to be
  isotropic.
\figposition{Fig.~\ref{AngXi}}

In order to describe the reactions in the second category,
following elementary cross sections and the decay widths are required.
\begin{itemize}
\item[(F)] Scalar and vector meson production:
	\bitem
	\item[]
		${\bar K} N \to (\phi, a_0, f_0) \Lambda$.
	\eitem
\item[(G)] Decay of scalar and vector mesons:\
	\bitem
	\item[]
		$(\phi, a_0, f_0)  \to K^-K^+$.\
	\eitem
\end{itemize}
Compared to the processes of (A)--(E), 
	the experimental information on
	the scalar and vector meson ($\phi/f_0/a_0$) production
	is very scarce.
Here, 
	$\phi/f_0/a_0$ production cross sections and angular distributions
	are taken from Ref.~\cite{DG}:
\begin{eqnarray}
\sigma(K^- p \to \phi  + \Lambda) &=&
       0.31531 p_f\exp(-1.45 p_f) \quad [\mb], \\
\sigma(K^- p \to f_0 + \Lambda) &=&
          1.51\sigma(K^- p \to \phi  + \Lambda),\\
\sigma(K^- p \to a_0 + \Lambda) &=&
	  1.38\sigma(K^- p \to \phi  + \Lambda),\\
{ d\sigma \over d\Omega} & \sim & \exp\left(1.8\cos\theta\right)\ ,
\end{eqnarray}
where
	$p_f$ is the relative momentum for the final state.
The angular distribution of the $\phi$ decay into ${\bar K} K$
	is assumed to be isotropic in its rest frame.
But we have checked that
	an anisotropic decay ($\cos^2\theta$ distribution)
	has no significant influence in our results.

For two-step processes,
   in the case of the \nuc{12}{C} target,
Iijima {\it et al.}~\cite{Iijima} found that
  the calculated cross section is 
   smaller than the observed one
   in an order of magnitude.
Moreover,
 Gobbi {\it et al.}~\cite{DG}
  also suggested that two-step processes can not reproduce the
  experimental cross section at low momentum region
  in the $K^+$ spectrum.
Note, however, that 
   the intermediate meson is limited to the pion,
    and they considered only for the \nuc{12}{C} target case.
We expect that
  inclusion of heavy mesons as $\eta$ mesons
  would strongly affect the final results
  especially in the case of heavier nucleus.
Therefore,
	we take account of two-step processes on all targets
	by including following processes:
\begin{itemize}
\item[(H)] Single strangeness exchange reactions:
	$$
	{\bar K} N \to 
        (\pi, \rho, \eta, \omega, \eta') (Y,Y^*)\ .
        $$
\item[(I)] Single strangeness creation reactions:
        $$
		(\pi, \rho, \eta, \omega, \eta')N  \to
		(K, K^*)  (Y,Y^*),\ \phi N   \ .
        $$
\item[(J)] Decay of resonances:\
	\bitem
	\item[]
		$\rho \to 2\pi$,\
		$\omega \to 3\pi$.
	\eitem
\end{itemize}
Here, 
  $Y=\{\Lambda,\Sigma$\},
  $Y^*=\{\Lambda(1405),\Lambda(1520),\Sigma(1385)\}$
  and $K^*=\{K^*(892)\}$.
  We treat explicitly the isospin of all these particles
  in our calculations.

We have fitted the experimental data as far as they are available.
The fitted function is partly taken from Refs.~\cite{Cug,Koch}
(see Appendix).
Other unknown cross sections in the second step (I),
	are estimated using the Breit-Wigner formula
	in the same way as Ref.~\cite{Brown,Sorge}.
Namely, processes of (I)
	are assumed to
	occur through $N^*,\Delta^*$ resonances
	and interferences of various resonances are ignored.
Since
	the energy is located just at the baryonic resonance region
	in the $(K^-,K^+)$ reaction
	with the incident momentum of $p_{K^-}=1.65$ GeV/c,
	this assumption is expected to be valid.
The explicit form is expressed as
\begin{equation}
\label{eq:bwformula}
\sigma(MB \to M'B')= {\pi \over k_{cm}^2}
                     \sum_{R}{(2J_R+1) \over (2S_M+1)(2S_B+1)}
                     {\mit\Gamma_R(MB)\mit\Gamma_R(M'B') \over
                      (\sqrt{s}-m_R)^2+\mit\Gamma_R(tot)^2/4}
\end{equation}
with the momentum dependent width
\begin{equation}
  {\mit\Gamma}_R(MB)=\left({p \over p_R}\right)^{(2\ell +1)}
       {1.2m_R/\srt \over 1 +0.2(p/p_R)^{2\ell}} {\mit\Gamma}^0_R(MB) \ ,
\end{equation}
where ${\mit\Gamma}^0_R(MB)$ denotes
       the partial width of the $R\to MB$ decay,
and 
 \begin{eqnarray}
  p &=&
   {\sqrt{(\srt-(m_B+m_M))(\srt-(m_B-m_M))} \over 2\srt} \ ,\\
  p_R &=&
    {\sqrt{(m_R-(m_B+m_M))(m_R-(m_B-m_M))} \over 2\srt} \ ,
 \end{eqnarray}
are the relative momenta between the meson and the baryon
 in the final state at a given c.m. energy
 and the resonance mass, respectively.
The summation $R$ in Eq.(\ref{eq:bwformula}) runs over resonances,
	as $N(1440)\sim N(2190), \Delta(1600)\sim\Delta(1950)$.
This formula is called as the generalized Breit-Wigner formula,
	which is obtained by neglecting
	$t$ channel contributions
	and
	the interference of the resonances.
Actual values for these parameters are 
 taken from the Particle Data Group~\cite{PG},
 as listed in Tables~\ref{bwD} and \ref{bwN}.

\figposition{Table.~\ref{bwD}}
\figposition{Table.~\ref{bwN}}

The formula Eq.~(\ref{eq:bwformula}) enables us to estimate
  experimentally inaccessible cross sections
  such as $\rho N \to \Lambda K$.
The $N^*$ resonances decaying into
   the $\omega N$ and
   $\eta' N$ branches
 are included to fit
    the experimental data of
       $\pi^- p  \to \omega n(\eta' n)$~\cite{HERA1}.
It has been shown that
  such meson resonances
  play an important role to product strangeness particles
  in heavy-ion collisions
  at AGS and SPS energies~\cite{Sorge}.
In fact,
 strangeness production cross sections
 via the meson resonances
  are found to be much larger
  than $\pi N$ cross sections.

Although $t$ channel component of $\bar{K} N$ interactions are
 considered to be large in (H),
 we also apply the resonance formula of Eq.(\ref{eq:bwformula})
   taking into account
   hyperon resonances
      $\Lambda(1520)\sim\Lambda(2350)$ and
      $\Sigma(1660)\sim\Sigma(2250)$
      in order to obtain the cross sections.
However,
 there is a lack of experimental information 
 about resonance parameters
 compared with $N^*$ and $\Delta^*$.
Therefore,
 we determined the experimentally unknown resonance parameters
   to fit the existing data~\cite{HERA2}
     such as $\Km p \to \eta \Lambda$.
 The parameters are shown in Tables~\ref{bwL} and \ref{bwS}.
 This procedure would be sufficient for this study
  because we need only ${\bar K} N$ interaction cross sections
     as the incoming states.
If we cannot obtain
 the cross sections and angular dependencies
   from the Breit-Wigner formula or from the experimental data,
   for example, $\sigma(\rho^- p \to \Sigma^0 K^0(892))$,
  we simply assume that
      these values are the same as $\pi N$ cross sections
       at the same c.m. energies.
For the total cross section of $\phi N$,
   we have used the value of 8.3mb,
	which was suggested 
	in a photoproduction experiment~\cite{phiN},
 while 16.0mb is predicted in
  the additive-quark model Ref.~\cite{Sorge}.

\figposition{Table.~\ref{bwL}}
\figposition{Table.~\ref{bwS}}

Note that
 the process $\phi N \to K \Lambda$
  is also able to produce the $K^+$ meson.
The cross section
  calculated from the OBE model of Ref.~\cite{Ko}
  amounts to about 5mb, as seen in  Fig.~\ref{fig:phin}.
Our INC calculation
	includes this process as a two-step process.

\figposition{Fig.~\ref{fig:phin}}

\section{Results and Discussions}\label{sec:results}

\subsection{$\Xi$ and $\Xi^*(1535)$ Productions}\label{sec:xi}

Let us now compare our results with the experimental data. 
At first,
  we focus on the $\Xi$ and $\Xi^*(1535)$ production processes.
In Fig.~\ref{KKxi},
   we display the INC spectra
      using both the matter density (solid lines) and 
                 the Fermi-type nuclear density (dashed lines)
   for $(K^-,K^+)$ reactions on
   \nuc{12}{C}, \nuc{27}{Al}, \nuc{63}{Cu},
   \nuc{107}{Ag} and  \nuc{208}{Pb} targets.
The RIA results with matter density are also presented
(dotted lines).
\figposition{Fig.~\ref{KKxi}}

We can see that
 the results with the Fermi-type density 
	well reproduce the experimental data at the high momentum region,
  while the results with the matter density
        underestimate the experimental data.
Our calculations with Fermi-type nuclear density
   for $\Xi$ and $\Xi^*$ production processes
   are consistent with DWIA of Ref.~\cite{Iijima},
	which was calculated
   by using the same nuclear density~\cite{nuc_data}.
It is also seen that
  the INC and the RIA results with matter density
 give almost the same results
  around the high momentum region of $K^+$,
  including the shape 
  and absolute value of the spectrum.
The difference can be found
           in the lower momentum region of $K^+$.
This is due to the different treatment of
 initial and final state interactions.
The eikonal approximation of Eq.~(\ref{zeff}) means that
  once the particle is scattered,
    the flux is assumed to be lost in the RIA model.
On the other side,
  the INC model treats the initial and final 
    state interactions explicitly,
    which ensure us that
      both loss and gain of the flux
      are automatically included.
For example,
 some parts of the scattered \Kp particles will
  be detected while the momentum becomes lower
  than that just after the production.

\subsection{Contributions of the scalar and vector meson decays}
\label{sec:phi}

Here we consider the effect through the scalar($a_0/f_0$)
  and the vector meson($\phi$) decays.
  The importance of this process were first suggested by 
        Gobbi {\it et al.}~\cite{DG}.
In Fig.~\ref{KKphi},
  we present
  the calculated results of the $K^+$ spectra (solid lines),
  taking into account the decay from 
    $\phi$, $a_0$ and $f_0$ mesons;
  the $\phi/f_0/a_0$ production
  and their decay into $K^-K^+$ significantly contribute
  to the $K^+$ spectrum at low momentum region.
However,
  these processes are not enough to reproduce
  the large bump in the experimental data.

\figposition{Fig.~\ref{KKphi}}

In order to test these results, 
  we also calculate the cross section of the above processes
  employing RIA.
The RIA results
  are shown in Fig.~\ref{KKphi} (dotted lines).
We find that
  INC and RIA calculations
  give us similar results
    for all targets.
However,
	our results are about 50\% smaller
	than the cross section in Ref.~\cite{DG}.
There are several probable reasons for this contradiction.
The one reason lies
   in the treatment of the Fermi motion in nuclei.
Since the $\phi/f_0/a_0$ mesons
  can be produced by
       a subthreshold particle production
   in the $(K^-,K^+)$ reaction
	  at $p_{K^-}= 1.65$ GeV/c,
  the effect of the Fermi motion is expected to be very important
      in this situations~\cite{Niita,Mosel}.
Thus we have checked the sensitivity of
      the treatment for the Fermi motion.
As seen in Fig.~\ref{KKalphi},
  we compare the results of following calculations
  for the \nuc{27}{Al} target
    with the matter density (matter),
    the local Fermi momentum
	with the Fermi type nuclear density (F+F)
and the matter Fermi momentum
	 with the Fermi type nuclear density (F+M).
It can be seen that
   the cross section in case (F+M)
   is enhanced by the factor 1.3.
  This treatment corresponds to the calculation in Ref.~\cite{DG}.

\figposition{Fig.~\ref{KKalphi}}

In addition, we mention that
  Gobbi et al.~\cite{DG} used an "effective" incident momentum
     which includes the effects of nucleon Fermi motion,
and regarded it as
   the incident momentum.
If this effective incident momentum is used
    in the original laboratory frame,
it gives a wrong kinematics.
In fact,
   the high momentum threshold of the $K^+$ particle 
 is artificially modified to be about 1.1 GeV/c in Ref.~\cite{DG},
while the actual threshold is about 0.8 GeV/c
  through $\phi/a_0/f_0$ decays.

\subsection{Two-Step Processes}\label{sec:two}
 
In Figs.~\ref{KKalla} and ~\ref{KKallb},
  we show the calculated $K^+$ spectra
	through two-step processes (dotted lines) on
   \nuc{12}{C}, \nuc{27}{Al},
   \nuc{63}{Cu}, \nuc{107}{Ag}
   and \nuc{208}{Pb} targets:
  Long dashed lines denote the contributions of
       the $\Xi$ and $\Xi^*$ production,
  dashed lines represent those from $\phi/a_0/f_0$ decay,
  and solid lines are the sum of all the processes
  considered in this paper.
It is clearly observed that
   two-step processes play an significant role 
   in $K^+$ yields,
   especially for heavier targets.
   This result is in contrast to Ref.~\cite{Iijima,DG}.
In addition, the calculated total spectra well reproduce
	the experimental data.

\figposition{Fig.~\ref{KKalla}}
\figposition{Fig.~\ref{KKallb}}

Before concluding that two-step processes are important,
	we should estimate theoretical ambiguities in
	the present calculation.
We consider that 
	the largest ambiguities come from the angular distributions
	of elementary processes.
Here
 we have used the experimental angular distributions if possible,
 for example,
  the $\Xi$, $\Xi^*$, $\phi$ productions
  and some part of two-step processes.
The experimentally unknown angular distributions of two-step processes
	such as $\rho N$
     are simply assumed to be
     the same dependence to that of $\pi N$ at same c.m.energy.
Therefore,
   it is necessary to estimate the sensitivity of the
    angular distributions.
In Fig.~\ref{KKtwostep},
  we represent the results of the two-step contributions
    assuming the isotropic angular distribution
    for all of the two-step reactions.
This choice decreases forward angle cross sections,
	but the difference of integrated yields is at most 30\%,
	and their contributions are still significant.
Due to the large uncertainty of
    the elementary angular distribution
     like $\rho N \to K Y$,
  more quantitative discussions are impossible at this stage.
However, we can conclude that
	a significant part of $K^+$ particles are
	produced though various two-step processes,
  since the use of isotropic angular distribution
    might give the minimum yields of $K^+$.

\figposition{Fig.~\ref{KKtwostep}}

Contrast to
   the previous works~\cite{Iijima,DG},
the reasons
  why two-step processes have large cross sections
  can be understood as follows.
The first reason is that
  we take account of
   the meson resonances
   $\rho,\eta,\omega$
   as a intermediate meson.
In addition to the increase of the number of processes 
	resulting in $K^+$ production, 
due to the large masses of meson resonances,
the strangeness productions through these mesons
  have a larger cross section
  than that through pions.
Note that
  the $\pi N$ reaction for strangeness productions
          is all endoergic one ($Q<0$),
	  whereas
	  $\rho$-, $\eta$- and  $\omega$-nucleon reactions
	   for strangeness productions
	  are exoergic one ($Q>0$).
Thus inclusion of these mesons is of importance.
Figure~\ref{KKtwostep2} displays
	the contributions of intermediate mesons.
It can be seen that
  $\pi$, $\eta$ and $\omega$
   mesons mainly contribute to
      the two-step cross sections.
It is of importance to see that
  the cross sections of meson resonances 
   are comparable to that of pions.
The reason of the small $\rho$-contribution
 is that
  the $\rho$ meson are likely to decay
  before its collision due to its large width.
  We find also that
  pions from the $\rho$ meson decay can
   contribute to yields of $K^+$.
We confirm that
    if we include only pions as the intermediate meson
       and neglect hyperon resonances,
    this calculation gives similar result to
    the estimation of Iijima {\it et al.}
    in the case of \nuc{12}{C} target:
  As expected from the small mass number,
  the contribution from the two-step processes
   is rather smaller than that for the heavier targets.

\figposition{Fig.~\ref{KKtwostep2}}

The second reason is related to the meson momentum region.
The mesons produced
     by $K^-$ at $p_{K^-}=1.65$GeV/c
 are 
 just in the baryon resonance region
 where the cross sections have the largest values.
Alternatively,
 at the higher momentum of $K^-$,
  the contribution of the $\phi$ meson and its decay
  becomes larger,
  and $K^+$ from the $\phi$ meson decay
  would dominate $K^+$ yields.

We can also consider
 the $\phi N \to K^+ \Lambda$ process.
Using the OBE model~\cite{Ko},
we can obtain the cross section about 5mb at most.
Although this process is taken into account in the INC calculation,
 it is found that
 the contribution is small.

\section{Summary and Conclusions}\label{sec:sumkk}

In this paper,
  we have analyzed \KK reactions
     within a INC model
     that is newly developed especially
     to include various inelastic channels
     in order to investigate
     the whole momentum region of the $K^+$ momentum.
The detailed comparison
 for the $K^+$ spectrum
      has been done
        between the INC and the RIA model calculations.
We have shown that the results with these two models reasonably agree
	with each other in one-step processes,
	and the calculated total spectra with INC
	well reproduce the experimental data on various targets.

Our results show that
  the momentum distribution of $K^+$
  can be explained 
  mainly by following mechanisms:
\begin{itemize}
\item[(1)] Direct-type reactions:
	\bitem
	\item[]
 	$K^- p \to K^+ \Xi^-$ , 
	\item[]
	$K^- p \to K^+ \Xi^{*-}(1530)$ .
	\eitem
\item[(2)] Decay of scalar/vector mesons:
	\bitem
	\item[]
	$K^- p \to \left\{
	  \begin{array}{c}
	  \phi \\
	  a_0  \\
	  f_0
	  \end{array}\right\}
	  \Lambda,
	  \quad 
	  \left\{
	  \begin{array}{c}
	  \phi \\
	  a_0  \\
	  f_0
	  \end{array}\right\}
	  \to K^-K^+$.
	\eitem
 \item[(3)] Two-step processes:
	\bitem
	\item[]
   $
    K^- N \to \left\{ 
    \begin{array}{c}
	Y    \\
	Y^*  
    \end{array} \right\}
	M, \quad\mbox{followed by}\quad
    MN \to \left\{
      \begin{array}{c}
     K \\
     K^*
     \end{array}\right\}
     \left\{
     \begin{array}{c}
     Y \\
     Y^*
     \end{array}\right\}
   $,

   where $M=\pi, \rho, \eta, \omega$.
	\eitem
 \end{itemize}
For the process (1) and the process (2),
 the consistent results are obtained
    between the INC and the RIA models.
Among two-step processes (3),
 it is important to include
  $\pi$, $\rho$, $\eta$, $\omega$ mesons,
    and hyperon and kaon resonances
    to reproduce the $K^+$ spectrum.
There are mainly two reasons of 
  the enhancement of the two-step contributions,
  contrary to the estimation
  in the previous works~\cite{Iijima,DG}.
The first one is the variety of intermediate mesons and final hyperons.
	It is clear that a number of the path
	   resulting in $K^+$ production 
	   is much larger,
	   because various mesons and hyperon resonances are included.
	In addition,
	  meson resonances have larger masses than pions, 
	and stored energies in masses
	    are released in second step reactions.
	Thus the $K^+$ production cross section
	 with these meson resonances
	 becomes larger than that of pions.
The second reason is related to the meson momentum region.
	The mesons produced in this reaction 
	is just in the baryon resonance region
	where the cross sections have the largest values.

In conclusion,
  the INC model can explain consistently the \Kp momentum spectrum
  from the $(K^-,K^+)$ reaction at $p_{K^-}=1.65$GeV/c
  on various targets.
Both the two-step strangeness exchange and production processes
with various intermediate mesons
      and
$\phi$, $a_0$ and $f_0$ productions
    and their decay into $K^+K^-$ of which
	was first quoted in Ref.~\cite{DG},
are necessary to reproduce the experimental data
	in the low momentum region.

There are still some theoretical ambiguities and problems
	in the present study.
One of them is, as already discussed in Sec.~\ref{sec:results}, 
	the angular distribution of elementary processes.
Another problem is that we have ignored the propagation of heavy baryon
	resonances
	 which are assumed to be formed in Eq.~(\ref{eq:bwformula});
	however,
	 the lifetimes of these heavy resonances are so small that
	they are expected to decay before interacting with other nucleons
	at around the normal nuclear density.
The third one is the branching ratios of heavy baryon resonances,
	especially decaying into $\omega N$.
Since the measurement of decaying into $\omega N$ is relatively difficult,
	the experimental information of the branching ratio is scarce.
Although we have fitted the $\omega$ production cross sections and
	determined the plausible branching ratios,  
	it may be problematic to assume this reaction occurs only in 
	$s$ channel.
These facts may be related to the overestimation of the $K^+$ spectra
	at very low momentum region.
	
There are some interesting problems suggested in this work.
First, we note that
	the \KK reaction can produce hyperon resonances $Y^*$
	in the nuclear medium.
This opens up the possibility of studying the properties
  of the hyperon resonances in nuclear matter.
Another interesting problem is
     to predict yields of
         double-$\Lambda$ hyperfragment production
	and to make their production mechanism clear. 
This is possible if we extend our cascade code
  including mean field effects~\cite{QMD1,QMD2,QMD3,QMD4,Maru1,Maru2,nQMD}.
In order to search for double-$\Lambda$ hypernucleus experimentally,
  the $(K^-,K^+)$ experiment
  is planned at AGS(E906)~\cite{Fukuda,E906} in the near future.
The production mechanism of hyperfragments
     may be different from
     the $K^+$ productions through
      direct $\Xi$,
      direct $\Xi^*$,
      $\phi$ decay
      and two-step processes.
The study in this line is now in progress.

\ack{
The authors are grateful to Dr. T. Iijima for supplying us
with the experimental data.
 The authors also thank Prof. K. Imai, Dr. H. Sorge and Prof. S.N. Yang for 
 useful suggestions,
 and all the members of the Nuclear Theory Group
 in Hokkaido University for great encouragements. 
This work was supported by the Japan Society for Promotion of Science,
 and in part by
 the Grant-in-Aid for Scientific Research (No.\ 06740193 and No.\ 07640365)
 from the Ministry of Education, Science and Culture, Japan.
}

\appendix
\section{Parameterizations of Cross Sections and Angular Distributions}

\subsection{Cross Sections}

We list up the cross sections parametrzied with
 the experimental data.
Some parameterizations are taken from
          Ref.~\cite{Cug,Tsushima,Koch}.
\begin{equation}
\sigma(K^- p \to \Lambda \eta')=
    {0.02(\srt-\sqrt{s_0})^{0.4} \over(\srt-2.20726)^2 + 0.0589288}
  + {0.000186093(\srt-\sqrt{s_0})\over(\srt-3.07199)^2 + 0.00652879}
\end{equation}
\begin{eqnarray}
& & \sigma(\pi^- p \to \Lambda K^0) =
  { 0.007665(\srt-1.613)^{0.1341} \over (\srt-1.72)^2+0.007826} \\
& & \sigma(\pi^- p \to \Sigma^- K^+) =
     { 0.009803(\srt-1.688)^{0.6021} \over 0.006583+(\srt-1.742)^2}
     +{0.006521(\srt-1.688)^{1.4728} \over 0.006248+(\srt-1.940)^2} \\
& & \sigma(\pi^- p \to \Sigma^0 K^0) =
     {0.05014(\srt-1.688)^{1.2878} \over 0.006455+(\srt-1.730)^2} \\
& & \sigma(\pi^- p \to \Lambda(1405) K^0) =
      {0.02655(\srt-2.018)^{0.5378} \over (\srt-2.0754)^2+0.1808} \\
& & \sigma(\pi^- p \to \Lambda(1520) K^0) =
   {0.00963763(\srt-1.833)^{1.68811}\over(\srt-2.21759)^2+0.0249613} \\
& & \sigma(\pi^- p \to \Lambda K^0(892)) 
    =  0.240985q\exp(-1.09397q),\\
& & \qquad\qquad \mbox{where}\ 
     q={\sqrt{(s-(M_{\Lambda}+0.636))(s-(M_{\Lambda}-0.636))}
	 \over 2\srt} \\
& & \sigma(\pi^- p \to \Sigma^-(1385) K^+) =
        {0.00163105(\srt-1.752) \over (\srt-2.1833)^2 + 0.0114982 } \\
& & \sigma(\pi^- p \to \Sigma^0(1385) K^0) =
       {0.0113016(\srt-1.752) \over (\srt-2.12476)^2 + 0.0636864} \\ 
& & \sigma(\pi^- p \to \Sigma^- K^+(892)) =
          0.123117q\exp(-1.50466q), \\
& & \qquad\qquad \mbox{where} \
   q={\sqrt{(s-(M_{\Sigma^-}+0.636))(s-(M_{\Sigma^-}-0.636))}
	  \over 2\srt } \\
& & \sigma(\pi^- p \to \Sigma^0 K^0(892)) =
          0.188256q\exp(-1.32449q), \\  
& & \qquad\qquad \mbox{where} \ 
 q={\sqrt{(s-(M_{\Sigma^0}+0.636))(s-(M_{\Sigma^0}-0.636))}
      \over 2\srt } \\
& & \sigma(\pi^- p \to n \phi) =
        {0.000986473(\srt-1.948)^{0.045244}
                   \over (\srt-2.09)^2 + 0.0290906 } \\
& & \sigma(\pi^+ p \to \Sigma^+ K^+) =
          {0.03591(\srt-1.688)^{0.9541}\over 0.01548+(\srt-1.890)^2}
         +{0.1594(\srt-1.688)^{0.01056} \over 0.9412+(\srt-3.000)^2} \\
& & \sigma(\pi^+ p \to \Sigma^+(1385) K^+) =
    {0.0191992(\srt-1.833)^{0.983665}
       \over(\srt-1.98968)^2 + 0.0253142} \\
& & \sigma(\pi^+ p \to \Sigma^+ K^+(892)) =
     {0.0361293(\srt-1.829)^{0.1}
       \over(\srt-2.37123)^2 + 0.177898} \\ 
& & \sigma(\pi^0 p \to \Sigma^0 K^+) =
  {0.003978(\srt-1.688)^{0.5848} \over 0.006670+(\srt-1.740)^2}
  +{0.04709(\srt-1.688)^{2.1650} \over 0.006358+(\srt-1.905)^2} \\
& & \sigma(\pi^0 p \to \Sigma^+ K^0) =
       {0.05014(\srt-1.688)^{1.2878} \over 0.006455+(\srt-1.730)^2}
\end{eqnarray}

\subsection{$\pi N$ Angular Distributions}

The angular distributions are fitted as the following function,
\begin{equation}
  {d\sigma \over d\Omega}
\sim  a\exp(\gamma_1\cos\theta)+b\exp(-\gamma_2\cos\theta)
+c(1-\cos^2\theta)
\end{equation}
at the c.m. energy $s$ in GeV.

\paragraph{1. $\pi^- p \to \Lambda K^0$}

\begin{eqnarray}
 c &=& 0,\\
 a &=& \left\{
  \begin{array}{lr}
      -5616.87+6491.43s-1859.92s^2,\quad
                        & s \leq 1.878, \\
    \displaystyle{0.0962773\over (s-1.89039)^2+0.00650021},
			& \quad  s > 1.878,
  \end{array}\right.\\
 b &=& \left\{
   \begin{array}{lr}
      0.0                              & s< 1.878,\\
     -155.986 + 240.621s - 123.717s^2+21.2028s^3,\quad
		                       & 1.878 \leq s<2.024,\nonumber\\
   -4.72881 + s^{2.20857},\quad        & 2.024 \leq s<2.097,\\
   -65.8086+64.8666s-15.8761s^2,\quad  & 2.097 \leq s< 2.35,\\
   0.0, & \quad s \geq 2.35,
  \end{array}\right.\\
\gamma_1 &=& \left\{
   \begin{array}{lr}
   -9.54489 + 6.20974s,&  \quad s<1.878,\\
   28.4659-33.8566s+10.3904s^2,& s \geq 1.878,
  \end{array}\right.\\
\gamma_2 &=& 83.5865-55.3861s+8.68325s^2.
\end{eqnarray}

\paragraph{2. $\pi^- p \to \Sigma^- K^+$}

\begin{eqnarray}
c &=& 0.0 \\
a &=& \left\{
  \begin{array}{lr}
     0.0049,& \quad s<1.978,\\
     0.0,   & \quad s \geq 1.978,
  \end{array}\right.\\
b &=&  \left\{
  \begin{array}{lr}
    0.3632,& \quad s<1.978,\\
    1.0   ,& \quad s \geq 1.978,
  \end{array}\right.\\
\gamma_1 &=& 4.32675,\\
\gamma_2 &=&  \left\{
  \begin{array}{lr}
    1.38911,& \quad s<1.978,\\
    378.699-361.502s+86.4954s^2,& \quad 1.978 \leq s<2.097,\\
    -8.44056+4.53549s & \quad s \geq 2.097.
  \end{array}\right.\\
\end{eqnarray}

\paragraph{3.$\pi^- p \to \Sigma^0 K^0$}

\begin{eqnarray}
      c &=& 0.0, \\
      a &=& 0.00537342, \\
      b &=& 0.391835, \\
      \gamma_1 &=& 5.17482, \\
      \gamma_2 &=& 0.582968.
\end{eqnarray}

\paragraph{4. $\pi^+ p \to \Sigma^+ K^+$}

\begin{eqnarray}
      a &=&  \left\{
  \begin{array}{lr}
        0.00806605, & \quad s< 1.9,\\
        0.000323596,& \quad s \geq 1.9,
  \end{array}\right.\\
      b &=&  \left\{
  \begin{array}{lr}
	10.0601,   & \quad s< 1.9,\\
        0.00018645,& \quad s \geq 1.9,
  \end{array}\right.\\
      c &=&  \left\{
  \begin{array}{lr}
	0.0,     & \quad s< 1.9 \\
        38.3724  & \quad s \geq 1.9,
  \end{array}\right.\\
      \gamma_1 &=&  \left\{
  \begin{array}{lr}
	 7.49409, &  \quad s < 1.9,\\
         13.3355, & \quad s \geq 1.9,
  \end{array}\right.\\
      \gamma_2 &=&  \left\{
  \begin{array}{lr}
	0.889499, & \quad s < 1.9,\\
        13.3355.  & \quad s \geq 1.9.
  \end{array}\right.\\
\end{eqnarray}

\newcommand{\NKb}{$N{\bar K}$}
\newcommand{\NKbs}{$N{\bar K(892)}$}
\newcommand{\No}{$N\omega$}
\newcommand{\Ned}{$N\eta'$}
\newcommand{\Lp}{$\Lambda \pi$}
\newcommand{\Le}{$\Lambda \eta$}
\newcommand{\Lo}{$\Lambda \omega$}
\newcommand{\Lpr}[1]{$\Lambda \hbox{(#1)}\pi$}
\newcommand{\Sp}{$\Sigma \pi$}
\newcommand{\Se}{$\Sigma \eta$}
\newcommand{\Spr}[1]{$\Sigma \hbox{(#1)}\pi$}
\newcommand{\Xk}{$\Xi K$}
\newcommand{\DK}{$\Delta K$}
\newcommand{\Nsp}{$N^* \pi$}
\newcommand{\Np}{$N \pi$}
\newcommand{\Dp}{$\Delta \pi$}
\newcommand{\Nr}{$N \rho$}
\newcommand{\Ne}{$N \eta$}
\newcommand{\Nsi}{$N \sigma$}
\newcommand{\Sk}{$\Sigma K$}
\newcommand{\Lk}{$\Lambda K$}
\begin{table}[htbp]
\caption[]{
  Branching ratios for $\Delta^*$ decay
	  }
  \label{bwD}
\begin{center}
\begin{tabular}{cccccccccc}\hline\hline
& $J^p$ & Mass & Width&\Nsp & \Np  & \Dp  & \Nr & \Sk&$N^*$ type\\\hline
$P_{33}$&${3\over2}^+$&1.60&0.35&0.15&0.175&0.55 &0.125 &0.00 &$N(1440)$\\
$S_{31}$ &${1\over2}^-$&1.62 &0.15&0.05&0.25 &0.50& 0.20&0.00 &$N(1440)$\\
$D_{33}$ &${3\over2}^-$&1.70 &0.30&0.00&0.20 &0.40& 0.3984&0.0016& \\
$S_{35}$ &${1\over2}^-$&1.90 &0.20&0.40&0.10 &0.05& 0.45&0.00 &$N(1440)$\\
$F_{35}$ &${5\over2}^+$&1.905&0.35&0.00&0.10 &0.14998& 0.75 & 0.0002 & \\
$P_{31}$ &${1\over2}^+$&1.91 &0.25&0.60&0.225&0.25& 0.15&0.00 &$N(1440)$\\
$P_{33}$ &${3\over2}^+$&1.92 &0.20&0.37&0.20 &0.40& 0.00 &0.03 &$N(1440)$\\
$D_{35}$ &${5\over2}^-$&1.93 &0.35&0.00&0.15 &0.00& 0.85&0.00 & \\
$F_{37}$ &${7\over2}^+$&1.95 &0.30&0.19&0.40 &0.30& 0.10&0.01 &$N(1689)$
\\\hline
\end{tabular}
\end{center}
\end{table}

\begin{table}[htbp]
\caption[]{
    Branching ratios for $N^*$ decay
	  }
  \label{bwN}
\begin{center}
\begin{tabular}{ccccccccccccc}\hline\hline
& $J^p$ &Mass & Width & \Nsi&\Np& \Dp&\Nr &\Ne & \Lk & \Sk &\No &\Ned
\\\hline
$P_{11}$&${1\over2}^+$&1.44&0.35&0.050&0.65&0.25&0.05
				    &0.00&0.00&0.00&0.00&0.00 \\
$D_{13}$&${3\over2}^-$&1.52 &0.12&0.049&0.55&0.225&0.175&0.001
			       &0.00 &0.00&0.00&0.00\\
$S_{11}$&${1\over2}^-$&1.535&0.15&0.050&0.45&0.05&0.05&0.40
			       &0.00&0.00&0.00&0.00\\
$S_{11}$&${1\over2}^-$&1.65 &0.15&0.050&0.70&0.05 &0.121
			      &0.01&0.06 &0.00&0.00&0.00\\
$D_{15}$&${5\over2}^-$&1.675&0.15&0.009&0.34&0.55 &0.09 
			      &0.01&0.001&0.00&0.00&0.00\\
$F_{15}$&${5\over2}^+$&1.68 &0.13&0.150&0.65&0.10 &0.10 
				&0.00&0.00 &0.00&0.00&0.00\\
$D_{13}$&${3\over2}^-$&1.70 &0.10&0.399&0.10&0.399&0.10 
				  &0.00&0.002&0.00&0.00&0.00\\
$P_{11}$&${1\over2}^+$&1.71 &0.10&0.015&0.15&0.15 &0.125
				  &0.30&0.20 &0.6 &0.00&0.00\\
$P_{13}$&${3\over2}^+$&1.172&0.15&0.100&0.15&0.10 &0.20 
				   &0.04&0.05 &0.06&0.30&0.00\\
$F_{13}$&${3\over2}^+$&1.90 &0.50&0.100&0.15&0.10 
				   &0.20 &0.04&0.05 &0.03&0.63&0.00\\
$P_{13}$&${3\over2}^+$&1.99 &0.35&0.100&0.15&0.10 &0.30 
				  &0.04&0.05 &0.03&0.50&0.03\\
$F_{17}$&${7\over2}^-$&2.19 &0.45&0.157&0.15&0.11 &0.30
				    &0.03&0.003&0.02&0.20&0.02\\
\hline
\end{tabular}
\end{center}
\end{table}

\begin{table}[htbp]
\caption[]{
    Branching ratios for $\Lambda^*$ decay
	  }
  \label{bwL}
\begin{center}
\begin{tabular}{ccccccccccc}\hline\hline
& $J^p$ &Mass & Width & \NKb & \Sp & \Le & \Xk& \Spr{1385}&\Lo& \NKbs 
\\\hline
$D_{03}$ &${3\over2}^-$& 1.52 & 0.015 & 0.46 & 0.43 & 0.00 
		      & 0.00 & 0.11  & 0.00 & 0.00\\
$P_{01}$ &${1\over2}^+$& 1.60 & 0.150 & 0.30 & 0.60 & 0.00 
                          & 0.00 & 0.10  & 0.00 & 0.00\\
$S_{01}$ &${1\over2}^-$& 1.67 & 0.035 & 0.20 & 0.45 & 0.35 
                          & 0.00 & 0.06  & 0.00 & 0.00\\
$D_{03}$ &${3\over2}^-$& 1.69 & 0.065 & 0.30 & 0.40 & 0.00 
			   & 0.00 & 0.45  & 0.00 & 0.00\\
$S_{01}$ &${1\over2}^-$& 1.80 & 0.300 & 0.40 & 0.20 & 0.00 
			  & 0.00 & 0.20  & 0.00 & 0.20\\
$P_{01}$ &${1\over2}^+$& 1.81 & 0.150 & 0.33 & 0.23 & 0.00 
			  & 0.00 & 0.01  & 0.00 & 0.43\\
$F_{05}$ &${5\over2}^+$& 1.82 & 0.080 & 0.62 & 0.14 & 0.04 
			  & 0.00 & 0.12  & 0.00 & 0.08\\
$D_{05}$ &${5\over2}^-$& 1.83 & 0.095 & 0.055& 0.74 & 0.02 
			  & 0.00 & 0.185 & 0.00 & 0.00\\
$P_{03}$ &${3\over2}^+$& 1.89 & 0.100 & 0.35 & 0.05 & 0.00 
			  & 0.00 & 0.10  & 0.35 & 0.15\\
$G_{07}$ &${7\over2}^-$& 2.02 & 0.200 & 0.41 & 0.23 & 0.00 
			  & 0.06 & 0.20  & 0.12 & 0.23\\
$G_{07}$ &${7\over2}^+$& 2.10 & 0.200 & 0.35 & 0.10 & 0.03 
			  & 0.03 & 0.18  & 0.001& 0.289\\
$F_{05}$ &${5\over2}^+$& 2.11 & 0.200 & 0.20 & 0.25 & 0.00 
			  & 0.00 & 0.03  & 0.00 & 0.56\\
$H_{09}$ &${9\over2}^+$& 2.35 & 0.150 & 0.24 & 0.20 & 0.00 
			  & 0.00 & 0.26  & 0.02 & 0.28\\
$D_{03}$ &${3\over2}^-$& 2.00 & 0.200 & 0.35 & 0.05 & 0.05 
			  & 0.05 & 0.05  & 0.15 & 0.30\\
\hline
\end{tabular}
\end{center}
\end{table}

\begin{table}[htbp]
\caption[]{
     Branching ratios for $\Sigma^*$ decay
	  }
  \label{bwS}
\begin{center}
\begin{tabular}{ccccccccccc}\hline\hline
& $J^p$ &Mass &Width &\NKb &\Lp &\Sp &\Se &\Xk &\Spr{1385} &\Lpr{1450}
\\\hline
$P_{11}$ &${1\over2}^+$&1.66&  0.10  & 0.30& 0.30 & 0.30 
        & 0.00& 0.00& 0.00& 0.00 \\
$D_{13}$ &${3\over2}^-$& 1.67&  0.06 & 0.13& 0.15 & 0.45
        & 0.00& 0.00& 0.09& 0.09 \\
$S_{11}$ &${1\over2}^-$& 1.75&  0.09 & 0.31& 0.05 & 0.08
        & 0.41& 0.00& 0.00& 0.00 \\
$D_{15}$ &${5\over2}^-$& 1.775& 0.12 & 0.43& 0.20 & 0.05
        & 0.00& 0.00& 0.12& 0.00 \\
$F_{15}$ &${5\over2}^+$& 1.915& 0.12 & 0.15& 0.20 & 0.15
	& 0.11& 0.04& 0.10& 0.00 \\
$D_{13}$ &${3\over2}^-$& 1.94 & 0.22 & 0.20& 0.10 & 0.10
	& 0.10& 0.00& 0.10& 0.20 \\
$F_{17}$ &${7\over2}^+$& 2.03 & 0.18 & 0.23& 0.23 & 0.10
	& 0.00& 0.02& 0.10& 0.00 \\
$F_{17}$ &${7\over2}^-$& 2.10 & 0.13 & 0.20& 0.20 & 0.00
	& 0.00& 0.00& 0.00& 0.00 \\
$D_{15}$ &${5\over2}^-$& 2.25 & 0.10 & 0.10& 0.12 & 0.10
	& 0.10& 0.00& 0.12& 0.00 \\\hline
\end{tabular}
\vspace*{3.6cm}
\begin{tabular}{ccccccccc}\hline
& &  &  &\Lpr{1520} &\DK &\NKbs & $\Xi (1530)K$ & $\Lambda\rho$ \\ \hline
 && &  & 0.00  & 0.05& 0.05 & 0.00 & 0.00\\
 && &  & 0.09  & 0.00& 0.00 & 0.00 & 0.00\\
 && &  & 0.10  & 0.05& 0.00 & 0.00 & 0.00\\
 && &  & 0.20  & 0.00& 0.00 & 0.00 & 0.00\\
 && &  & 0.00  & 0.13& 0.12 & 0.00 & 0.00\\
 && &  & 0.00  & 0.20& 0.05 & 0.00 & 0.00\\
 && &  & 0.10  & 0.20& 0.08 & 0.00 & 0.00\\
 && &  & 0.00  & 0.38& 0.10 & 0.00 & 0.12\\
 && &  & 0.00  & 0.24& 0.12 & 0.10 & 0.00\\ \hline
\end{tabular}
\end{center}
\end{table}

\newcommand{\getepss}[2]{\epsfxsize=#1in\epsfbox{#2}}
\newcommand{\getEPS}[1]{\epsfxsize=4in\epsfbox{#1}\vspace*{-74pt}}
\newcommand{\FIGN}[2]{%
\newpage{#2}\vfill\noindent{%
   \bf Y.Nara,A.Ohnishi,T.Harada and A.Engel: Fig.#1}}
\newcommand\putfigg[1]{%
\def\FIG{#1.ps}%
\centerline{\hbox{\vbox{%
\noindent%
\getEPS{RIA-Fig/kkc/\FIG}\\
\getEPS{RIA-Fig/kkal/\FIG}\\
\getEPS{RIA-Fig/kkcu/\FIG}\\
\getEPS{RIA-Fig/kkag/\FIG}\\
\getEPS{RIA-Fig/kkpb/\FIG}%
}}}}

\newcommand\geteps[2]{%
\centerline{\hbox{\psfig{figure=#2,height=#1in}}}
}
\def\FIGKKa{%
\begin{figure}
\getepss{5}{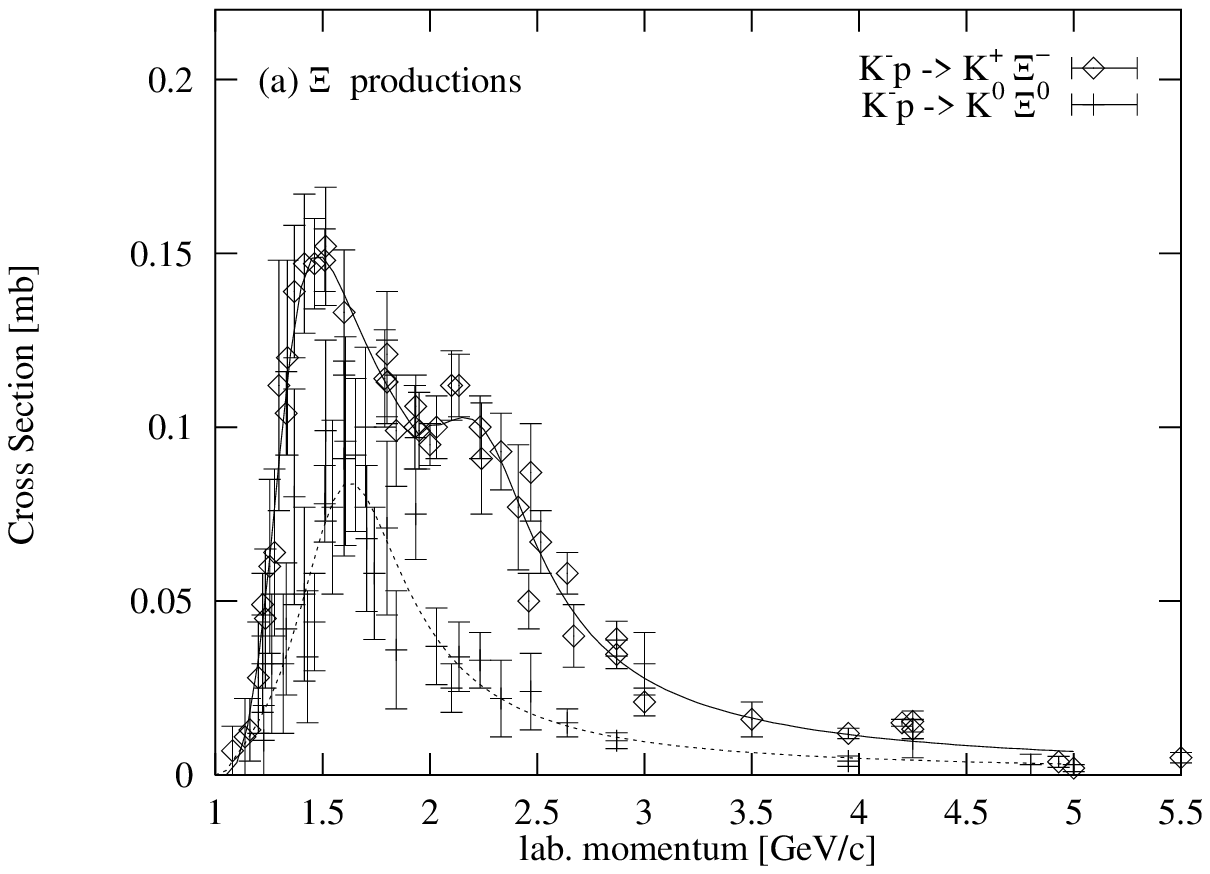}\getepss{5}{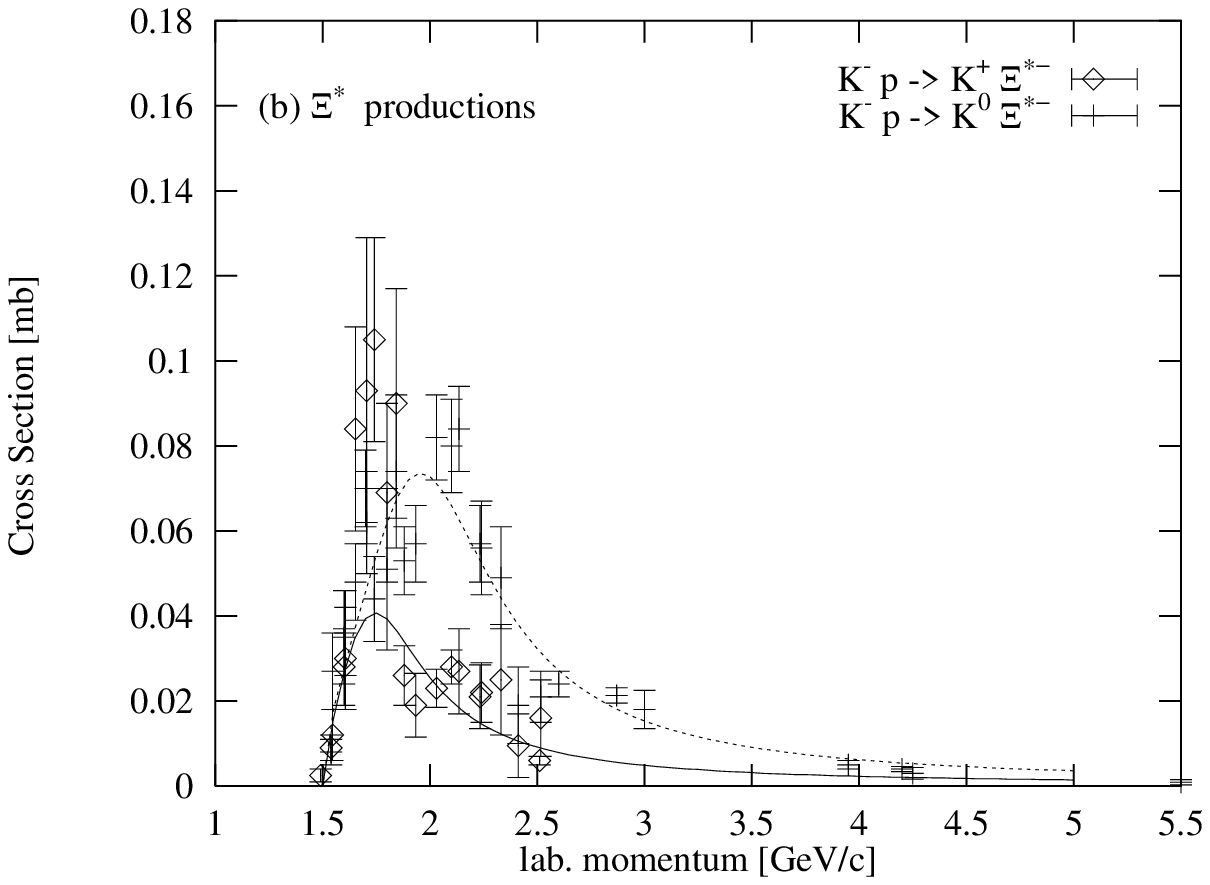}
\caption[]{
      Parameterizations of (a)$\Xi$ and (b)$\Xi^*$
      production cross sections with
      experimental data~\cite{Angxi}.
          }
      \label{CrossXi}
\end{figure}
}
\def\FIGKKb{%
\begin{figure}
\getepss{5}{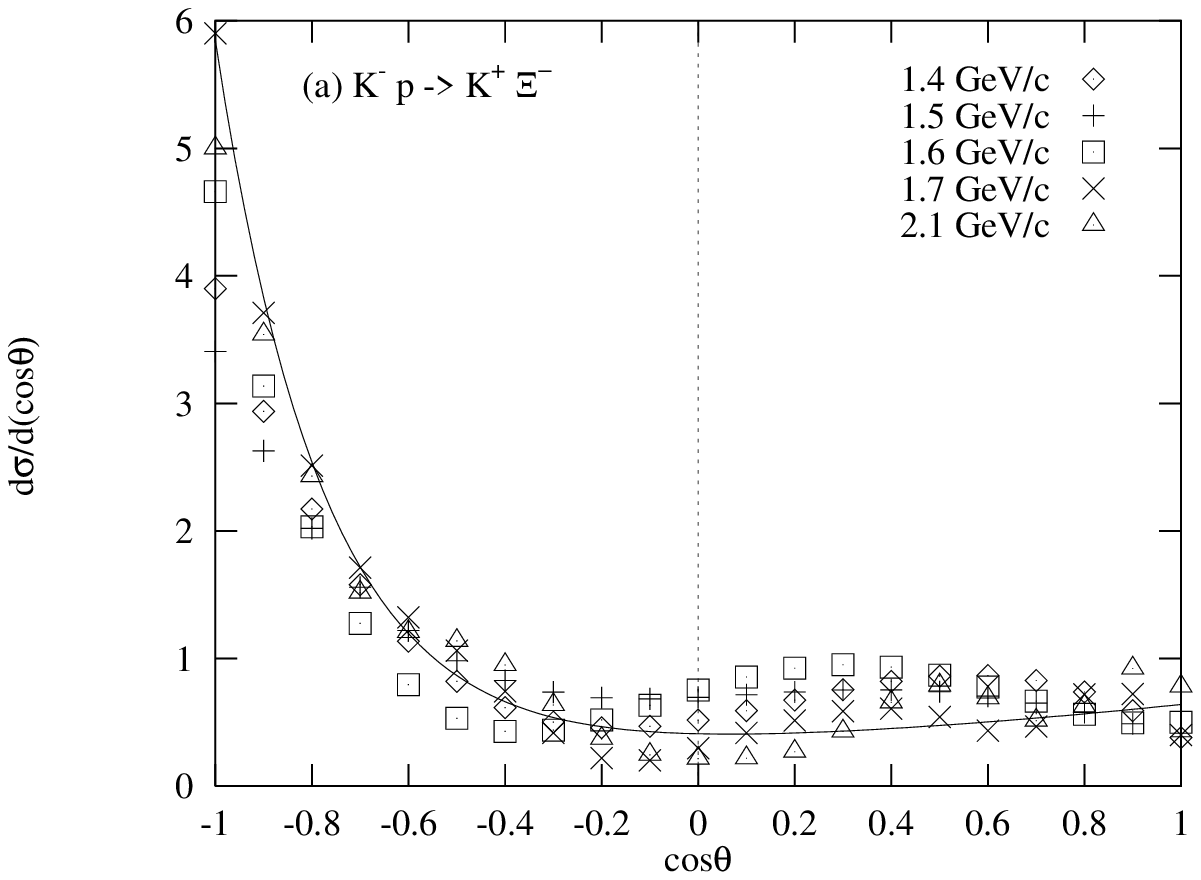}\getepss{5}{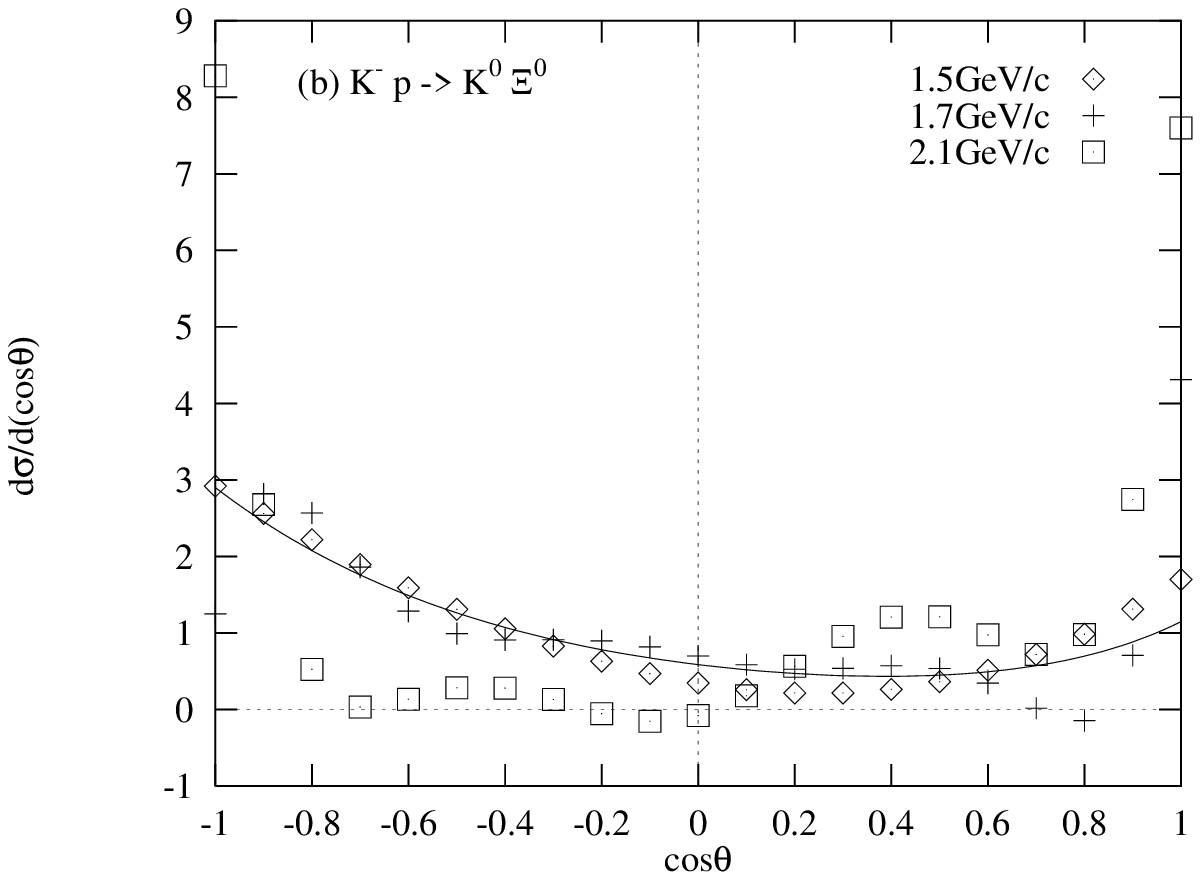}
\caption[]
      {
           Angular distributions of
           (a) $K^- p \to K^+ \Xi^-$
           and (b) $K^- p \to K^0 \Xi^0$.
	   The solid curve denotes for the angular distribution
           parameterized at the laboratory $K^-$ momentum of
	   $p_{K^-}=$ 1.7GeV/c.
      }
   \label{AngXi}
\end{figure}
}
\def\FIGphin{%
\begin{figure}
\geteps{5}{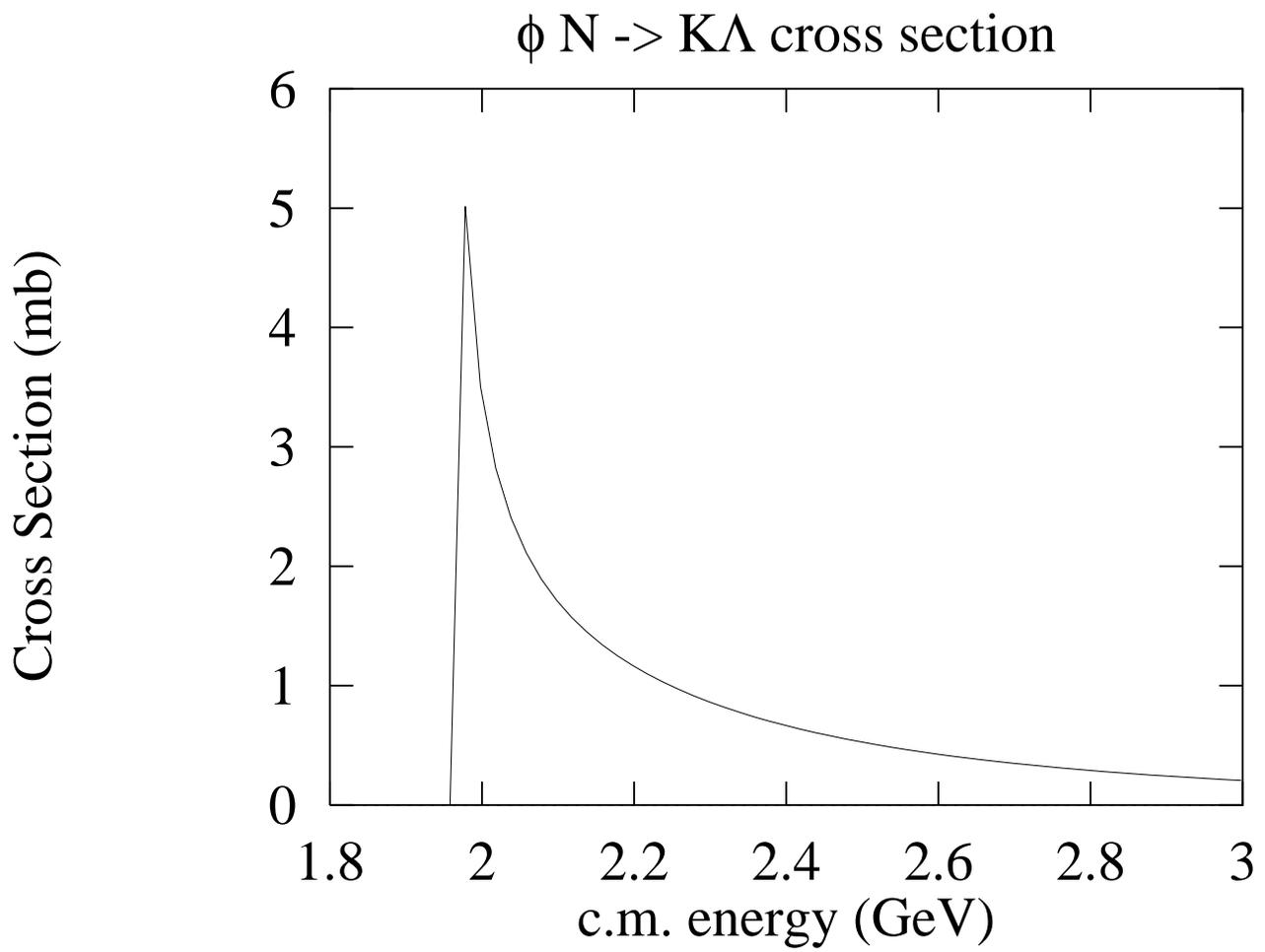}
\caption[]
      {
   $\phi N \to K\Lambda$ cross section
   predicted from OBE model of Ref.\cite{Ko}.
     }
   \label{fig:phin}
\end{figure}
}
\def\FIGKKc{%
\begin{figure}
\geteps{8}{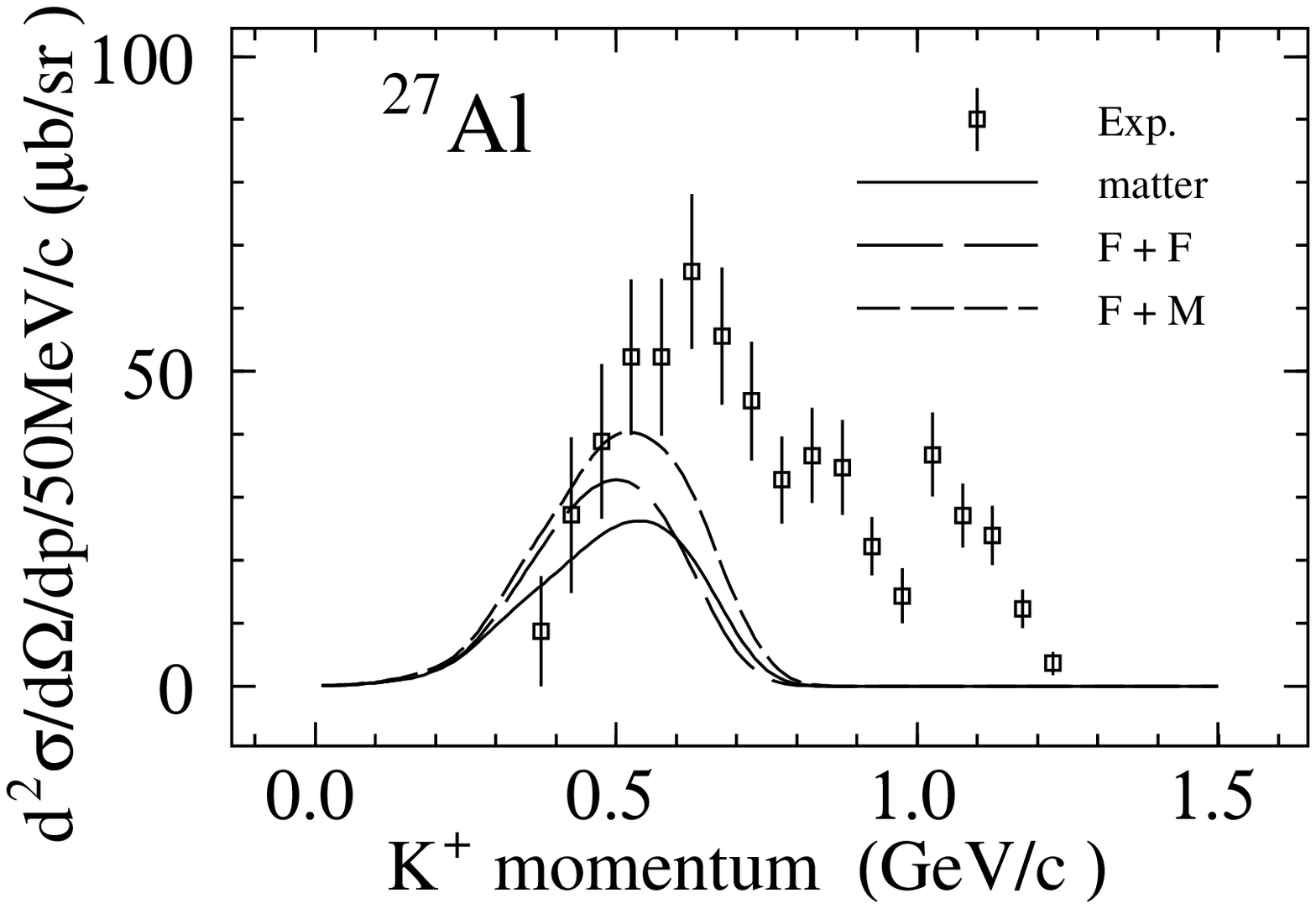}
\vspace*{-4cm}
\caption[]
      {
   \KK double differential cross sections on
   \nuc{27}{Al} target at $p_{K^+}=1.65$ GeV/c.
   The INC calculations only contain $\phi$, $a_0$ and $f_0$ productions.
   These spectra are calculated by using
   matter density (solid line),
   the Fermi density with the local Fermi momentum (F+F) (long dashed line)
   and
   the Fermi density with the matter momentum (F+M) (dashed line).
     }
   \label{KKalphi}
\end{figure}
}
\def\FIGKKriaA{%
\begin{figure}
\geteps{8}{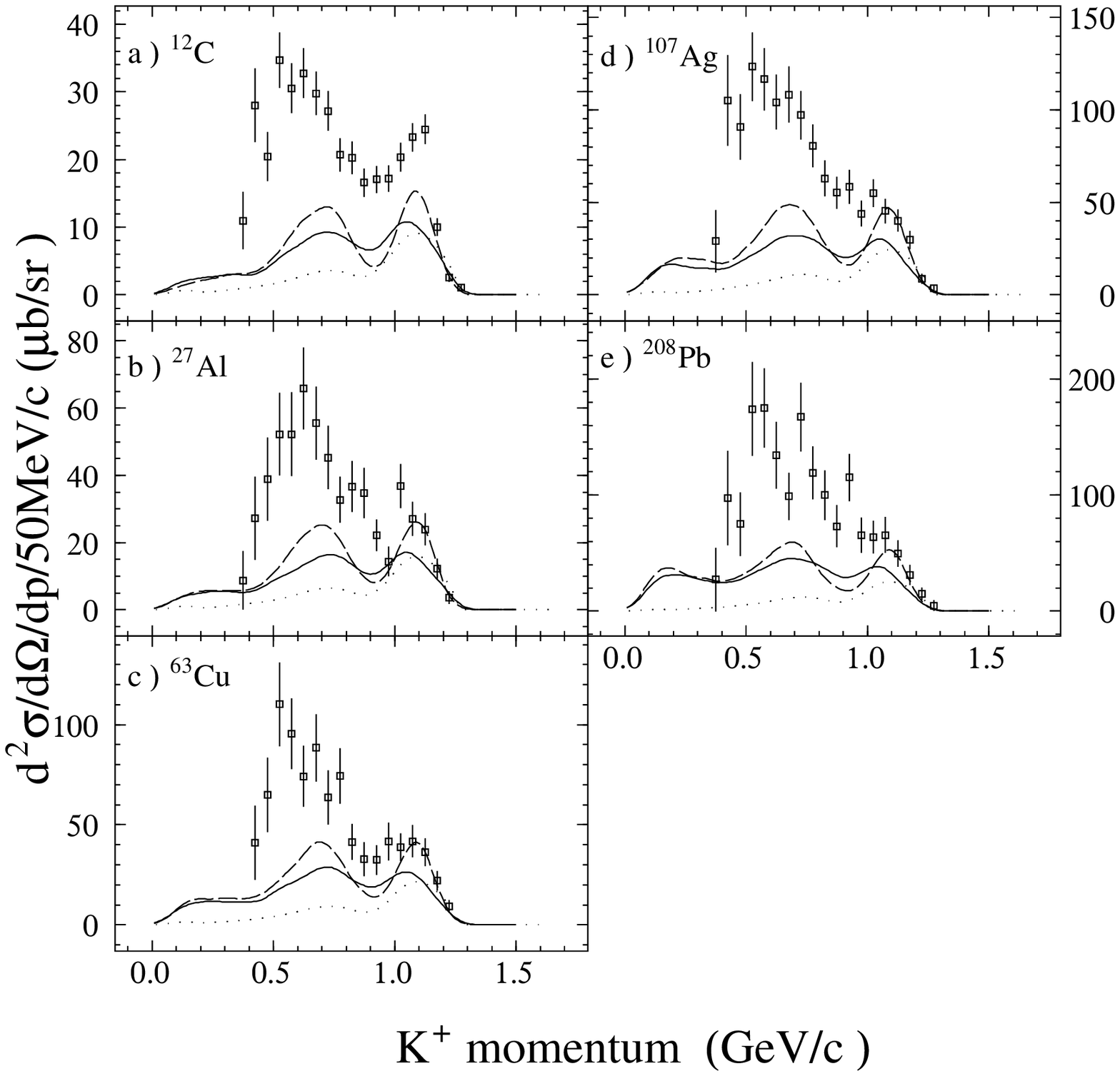}
\vspace*{-4cm}
\caption[]
      {
 	Comparison of the calculated results for the one-step processes
 	($\Xi^-$ and $\Xi^{*-}$ production) with the experimental data.
 	Solid and dashed lines show
 	the INC results with 
 	the matter density 
 	and with the Fermi-type density + local Fermi momentum,
 	respectively.
 	Dotted lines show the RIA results with the matter density.
 	\label{KKxi}
}
\end{figure}
}
\def\FIGKKriaB{%
\begin{figure}
\geteps{8}{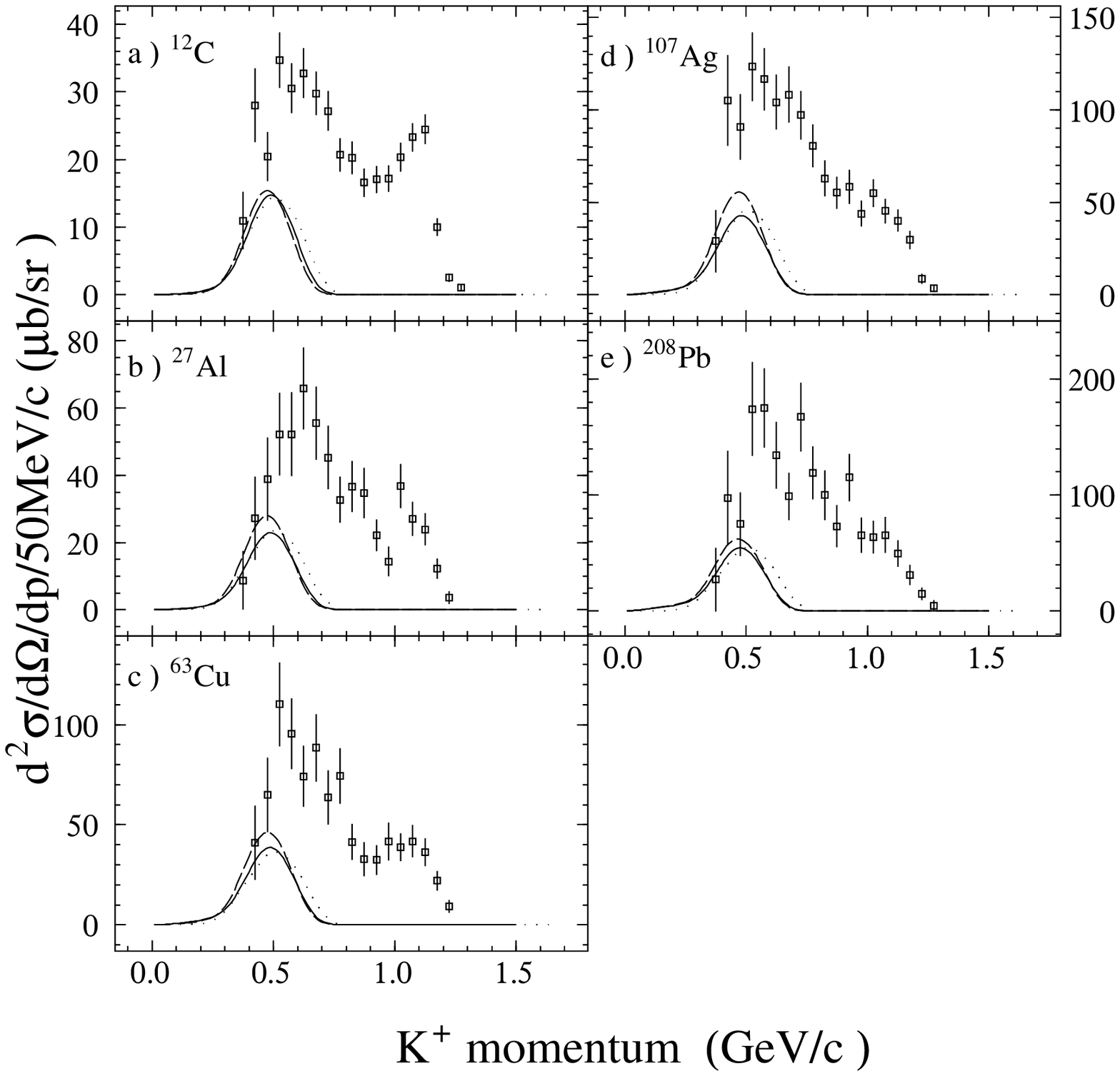}
\vspace*{-3cm}
\caption[]
      {
	Comparison of the calculated results
	for the meson-decay($\phi+a_0+f_0$) contributions
	to $K^+$ momentum spectra with the experimental data.
	Solid and dashed lines show the INC results with
	the matter density and the Fermi-type density,
	respectively.
 	Dotted lines show the RIA results with the matter density.
	\label{KKphi}
}
\end{figure}
}
\def\FIGKKf{%
\begin{figure}
\geteps{8}{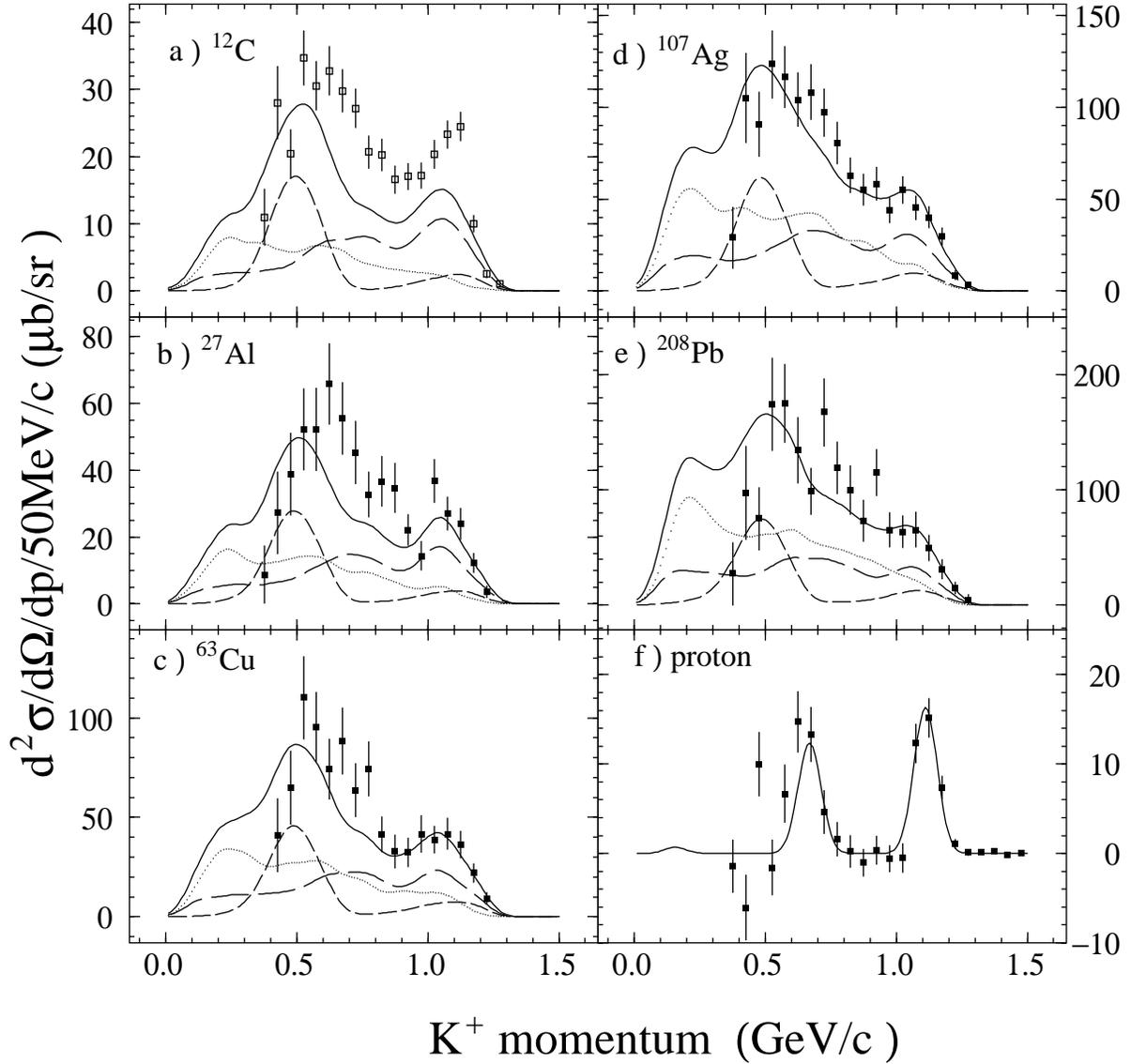}
\vspace*{-4cm}
\caption[]
      {
Calculated momentum spectra of $K^+$ 
 for \nuc{12}{C}, \nuc{27}{Al}, \nuc{63}{Cu}, \nuc{107}{Ag}, \nuc{208}{Pb}
 and proton targets
 at $p_{K^-}=1.65$ GeV/c using the INC model.
The matter density is used.
The squares represent the data of Iijima {\it et al}.~\cite{Iijima}.
The contributions of $\Xi$ and $\Xi(1535)$ productions
  are represented by long dashed lines.
  The dashed and dotted lines correspond to
  the contributions of $\phi$, $a_0$ and $f_0$ productions
  and of two-step processes, respectively.
Solid lines denote the results of the total spectrum.
   \label{KKalla}
}
\end{figure}
}
\def\FIGKKg{%
\begin{figure}
\geteps{8}{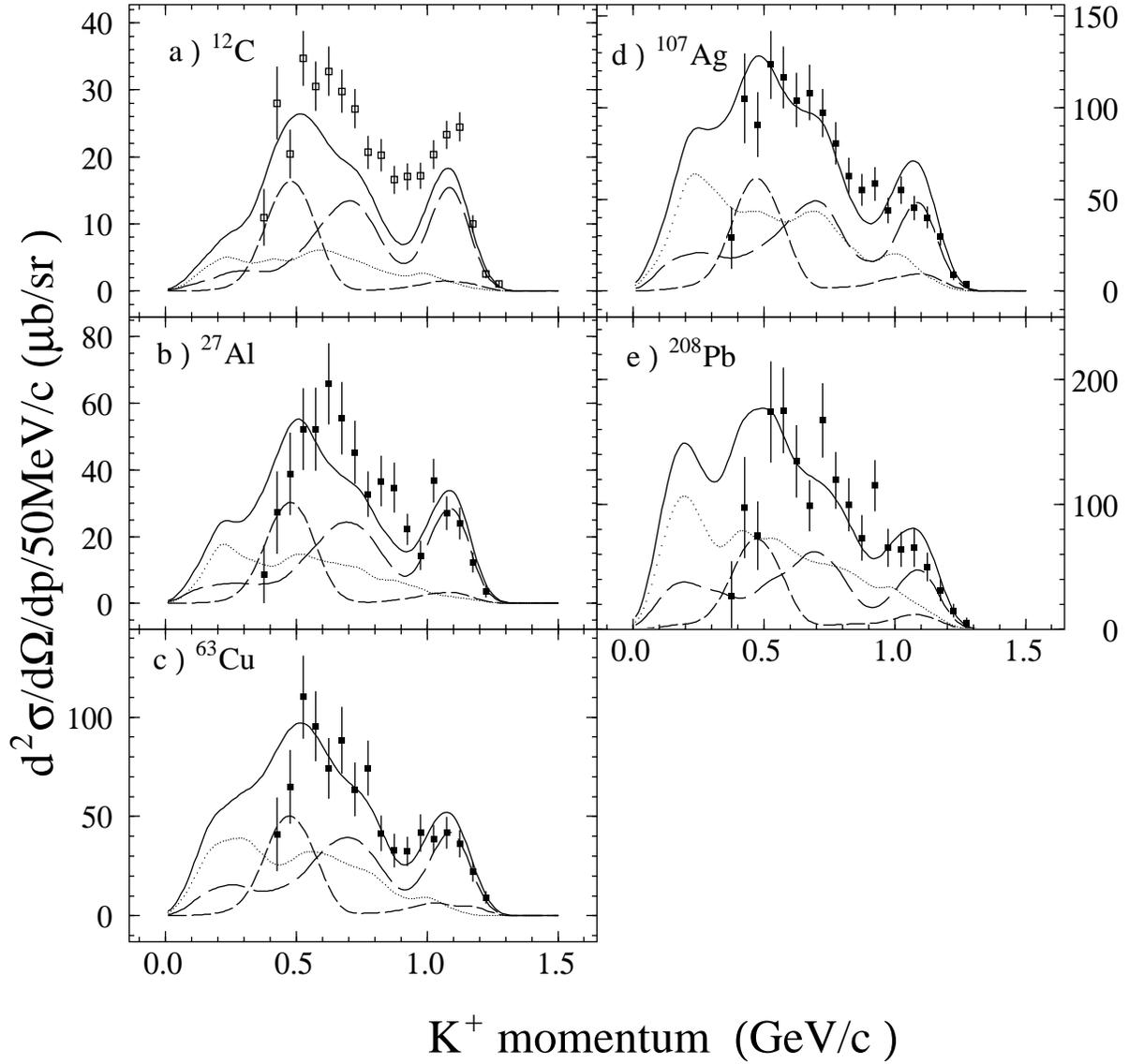}
\vspace*{-4cm}
\caption[]
      {
   Calculated momentum spectra of $K^+$
      by using the INC model
   for various targets at $p_{K^-}=1.65$ GeV/c with
      the Fermi type density for targets.
   The notation is the same as fig~\ref{KKalla}.
   \label{KKallb}
}
\end{figure}
}
\def\FIGKKd{%
\begin{figure}
\geteps{8}{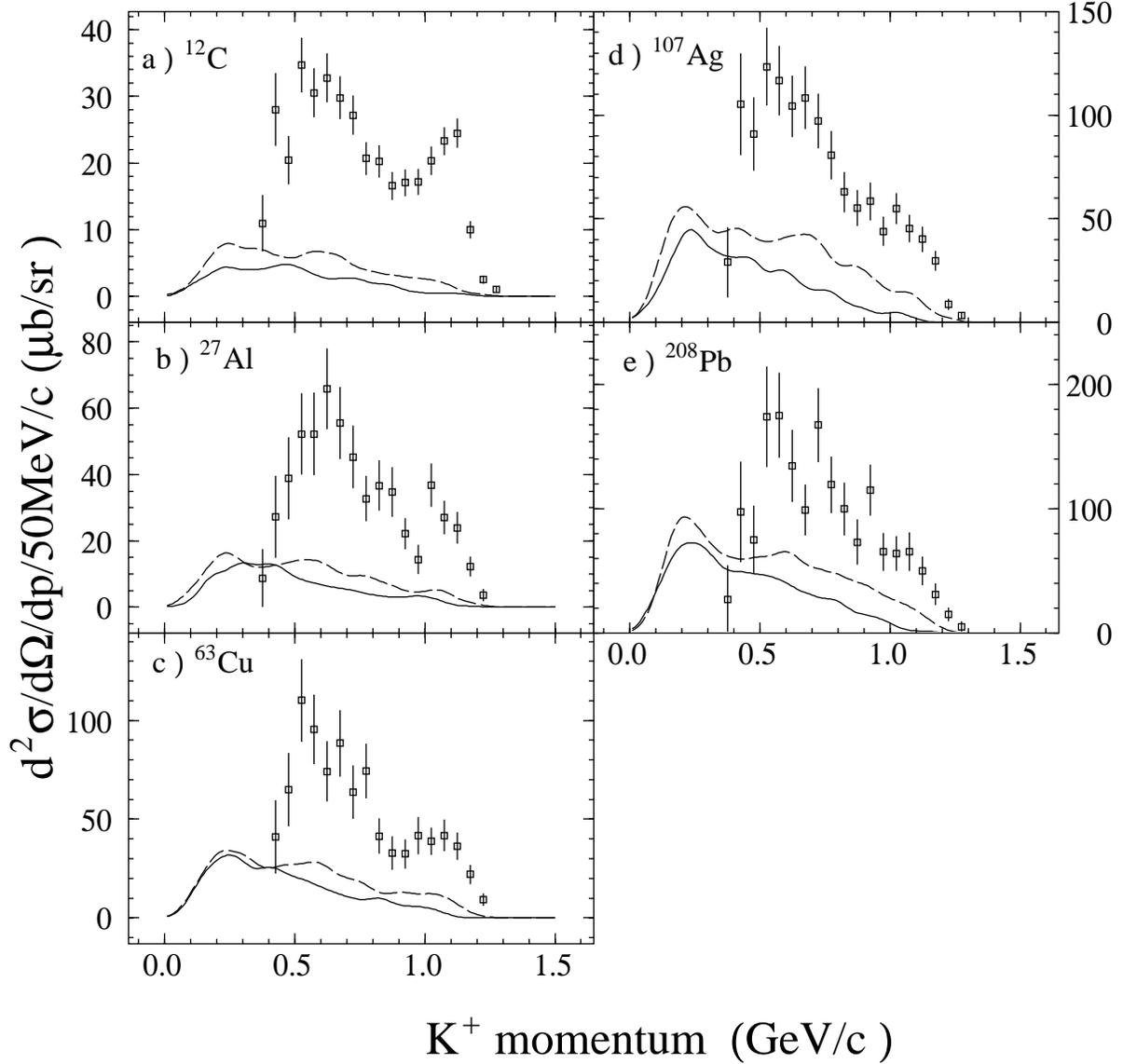}
\vspace*{-4cm}
\caption[]
      {
   \KK double differential cross sections
   for only the two-step processes in the INC model,
   together with the experimental data of Ref.~\cite{Iijima}.
   Solid lines correspond to the results
   with isotropic angular distribution
   for the $MN\to K(K^*)Y(Y^*)$, $M=\pi,\rho,\eta,\omega$ and $\eta'$.
   Dashed lines represent
     the results in which
      resonance-$N$ angular distributions are equated to
      those of $\pi N$ scattering.
   \label{KKtwostep}
}
\end{figure}
}
\def\FIGKKe{%
\begin{figure}
\geteps{8}{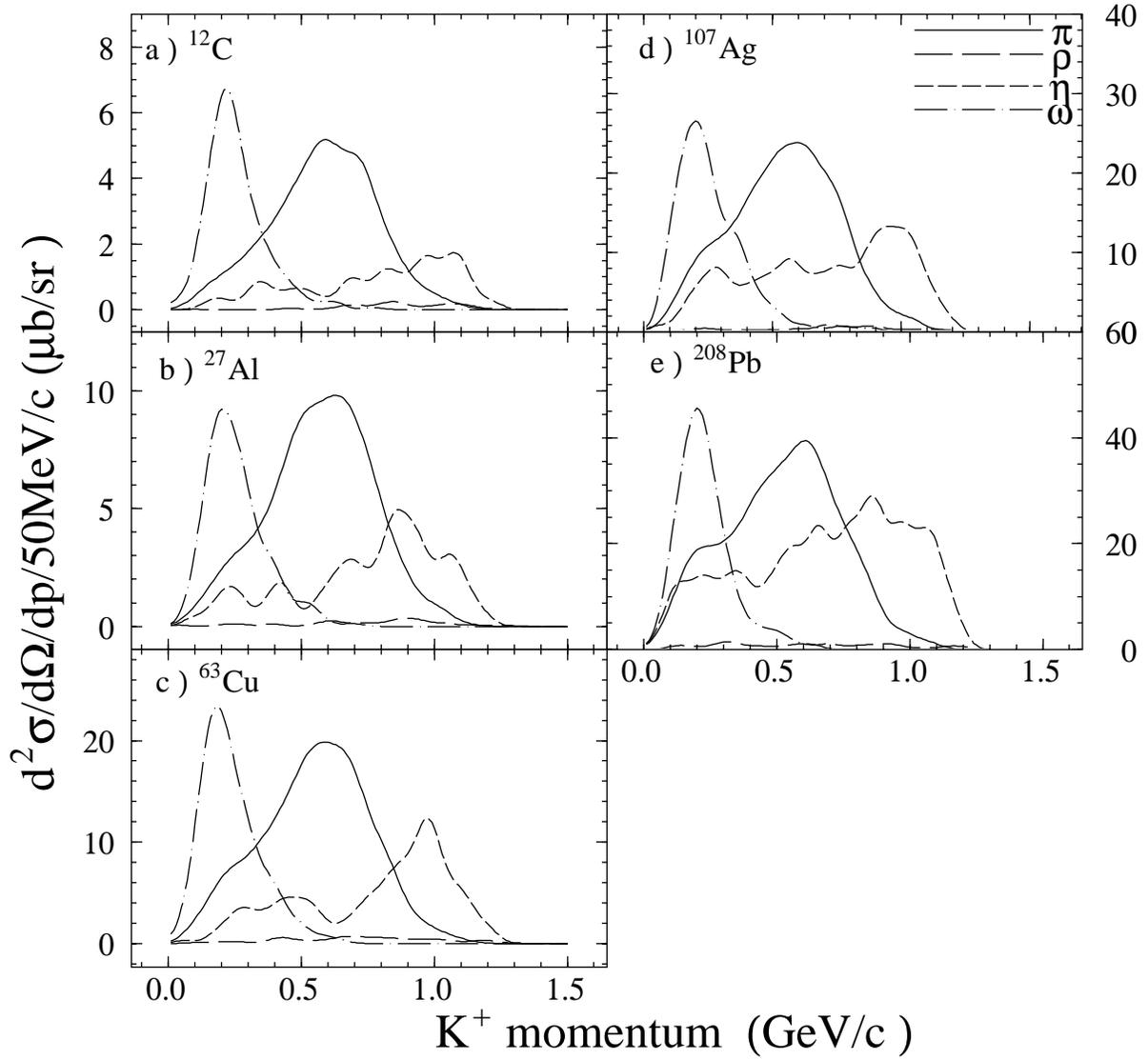}
\vspace*{-4cm}
\caption[]
      {
    Decomposition of the two-step processes
    calculated with the INC model.
    The contribution from
     $\pi$, $\rho$, $\eta$ and $\omega$ intermediate mesons
    are represented by solid,
    long dashed,
    dashed
    and dashed-dotted lines, respectively.
   \label{KKtwostep2}
}
\end{figure}
}

\FIGKKa     
\FIGKKb     
\FIGphin    
\FIGKKriaA  
\FIGKKriaB  
\FIGKKc     
\FIGKKf     
\FIGKKg     
\FIGKKd     
\FIGKKe     

\end{document}